\documentclass{aa}  
\usepackage{ae,aecompl}
\usepackage{adjustbox}
\let\linenumbers\nolinenumbers
\usepackage{graphicx}
\usepackage{txfonts}
\usepackage{hyperref}
\usepackage{xcolor}
\usepackage{diagbox}
\usepackage{orcidlink}
\usepackage[percent]{overpic}

\hypersetup{colorlinks,linkcolor={blue},citecolor={blue},urlcolor={blue}}  
\usepackage{graphicx}
\usepackage{epsf,palatino,url,wrapfig,array,setspace,amsmath,amssymb,fancyhdr,multirow,lscape,appendix,rotating}
\usepackage{graphics,epsfig,txfonts}
\usepackage{threeparttable}
\usepackage{xspace}

\newcommand{\swift}{\emph{Swift}\xspace}
\newcommand{\fermi}{\emph{Fermi}\xspace}

\newcommand{\code}{{\tt LeHaMoC}\xspace}

\begin{document}

\title{Time-resolved leptonic modeling of the prompt emission of GRB~211211A}
\authorrunning{Petropoulou et al.}
\titlerunning{Time-resolved modeling of GRB~211211A}
\author{Maria Petropoulou\orcidlink{0000-0001-6640-0179
}\inst{1}, 
Maria Gkoni\orcidlink{0009-0008-7499-9693}\inst{1}, 
Konstantinos Xyloportas\orcidlink{0009-0000-4836-2105}\inst{1,2}, 
Stamatios I. Stathopoulos\orcidlink{0000-0002-1344-3754}\inst{3}, 
Georgios Vasilopoulos\orcidlink{0000-0003-3902-3915}\inst{1}, 
Apostolos Mastichiadis\orcidlink{0000-0001-5217-4801}\inst{1}}  
\institute{Department of Physics, National and Kapodistrian University of Athens, University Campus Zografos, GR 15784, Athens, Greece
\email{mpetropo@phys.uoa.gr}
  \and
University Observatory, Faculty of Physics, Ludwig-Maximilians-Universität München, Scheinerstr. 1, 81679 Munich, Germany
\and 
Deutsches Elektronen-Synchrotron DESY, Platanenallee 6, 15738 Zeuthen, Germany
\email{stamatios.ilias.stathopoulos@desy.de}
   }
 
  \abstract
   {GRB~211211A is a long duration gamma-ray burst with a compact object merger origin. Due to its proximity to Earth and its minute-long duration, it provides one of the best temporally resolved GRB prompt emission spectra to date. }
   {In this work, we model the time-resolved prompt-emission spectra of GRB~211211A within a leptonic radiation framework. Our goal is to infer the physical properties of the emitting region, study the temporal evolution of the radiating particle distribution, and make predictions for prompt emission at TeV energies.}
   {We perform Markov Chain Monte Carlo fitting of the time-resolved numerical spectral energy distribution (SED) models computed with the time-dependent non-thermal radiation code \code. Our calculations include synchrotron emission and self-absorption, inverse Compton scattering including cooling in the Klein-Nishina regime, and photon--photon pair production.}
   {We find that the prompt emission of GRB~211211A between 10 keV and 10 MeV can be successfully reproduced by synchrotron radiation from a population of relativistic electrons. The spectral evolution during the first minute of the burst reflects different physical conditions in the emitting region. Our best-fit models favor fast-cooling solutions for the first 8~s, followed by a transition to slow-cooling solutions at later times. The accompanying synchrotron self-Compton emission extends to TeV energies, with predicted fluxes that would be detectable by CTAO for a burst similar to GRB~211211A, provided a sufficiently rapid response to a \fermi-GBM trigger or if the burst occurs within the CTAO field of view. }
   {The observed short variability of this burst requires very high Doppler factors ($\sim 1000-2500$) throughout the burst evolution. Such extreme Doppler factors are difficult to reconcile with the jet Lorentz factor inferred from afterglow modeling unless the prompt-emitting regions are themselves moving relativistically with respect to the jet plasma.}
   
   \maketitle
%

\section{Introduction} \label{sec:intro}
Gamma-Ray Bursts (GRBs) are the most luminous explosions in the Universe, exhibiting $\gamma$-ray peak (isotropic) luminosities  $\sim  10^{51}-10^{53}~\rm erg \, s^{-1}$ and broadband non-thermal photon emission that typically peaks between 100~keV and 1~MeV \citep[see][for a recent overview]{2022Galax..10...38B}. GRBs are believed to be associated with catastrophic events on stellar-mass scales, which is suggested by their energetics and rapidly varying emission on time scales ranging from milliseconds up to several seconds \citep[see, for example, Chapter 11 in][]{physics_grbs_11}.

Historically, GRBs have been classified into long or short events depending on the duration of the prompt $\gamma$-ray emission \citep{1993_Kouveliotou}. This duration is typically quantified through the time interval encompassing 5\% - 95\% of the cumulative photon count, known as $T_{90}$. Being an observational quantity, its measured value depends on detector characteristics, such as its energy range and sensitivity \citep[e.g.,][]{Bromberg_2012, Qin_2013, Moss_2026}. 

These two GRB classes have long been interpreted as reflecting distinct progenitor systems, with short-duration bursts ($T_{90} < 2$~s)  typically associated with compact-object mergers \citep[e.g.,][]{Eichler_1989,Narayan_1992,Abbott_2017} and long-duration bursts ($T_{90} > 2$~s) with the collapse of dying massive stars \citep[e.g.,][]{MacFadyen_1999, Woosley_Heger_2006}. Additional diagnostics, such as spectral evolution, host-galaxy properties \citep[e.g.,][]{Bloom_2002, Fong_2013}, and the presence of an associated supernova or kilonova \citep[e.g.,][]{Li_1998, Woosley_Bloom_2006, Metzger_2010, Tanvir_2013}, can further strengthen the identification of the progenitor system. Although it has long been recognized that the duration-based classification is not absolute, recent observations of lower-energy electromagnetic counterparts to specific GRBs made this ambiguity particularly evident: GRB~200826A \citep{Ahumada_2021, Rossi_2022}, GRB~211211A \citep{Rastinejad_2022, 2022_Troja}, GRB~230307A \citep{2026longGRBs_CompactMergers}.

GRB~211211A is classified as a long-duration burst based on its $T_{90}$ value, which was determined to be $51.37 \pm 0.80$~s in the \swift-BAT 15$-$350~keV energy band \citep{SwiftGRB211211A} and $34.3 \pm 0.6$~s in the 50$-$300~keV band \citep{Veres_2023}. Despite its duration, it presents strong indications of a compact object merger origin, with optical and near-infrared observations exhibiting a striking similarity with kilonova AT2017gfo \citep{Rastinejad_2022, 2022_Troja}. The isotropic-equivalent $\gamma$-ray energy of GRB~211211A calculated in the 1 keV - 10 MeV range is $E_{\rm iso} \approx 1.3 \cdot 10^{52}$~erg \citep{Veres_2023} is modest compared to the most energetic GRBs, which can reach $E_{\rm iso}>10^{54}$~erg \citep{2017_GRBmaxenergy}. Nonetheless, GRB~211211A is one of the brightest GRBs among the merger and collapsar populations of bursts detected by the Fermi-GBM \citep{Veres_2023}. 
No signal was detected by the Large Area Telescope (LAT) on board Fermi during the prompt emission phase, because the incident angle was greater than $65^\circ$ from the LAT boresight until $\sim800$~s after the GBM trigger time \citep{2021GCN.31210....1M, Mei_2022}. Transient-like GeV emission was detected by the LAT 1000~s after the main burst, which was associated with the interactions of the jet with the circumburst medium and external radiation fields \citep{Mei_2022}.

A weak precursor event, lasting $\sim0.15$~s \citep{2022_Troja, Veres_2023}, is associated with the prompt emission of GRB~211211A and was detected by both BAT and GBM. \citet{2024_Xiao} reported evidence for quasi-periodic oscillations (QPOs) during this precursor, a result that was subsequently supported by \citet{Lamb_2025}. These QPOs have been interpreted as possible signatures of a catastrophic flare accompanied with magnetoelastic or crustal oscillations of a magnetar formed after a binary compact merger.

The prompt emission of GRB~211211A also displays a bright, multi-peaked main burst and a highly time-variable, long-duration emission episode. Throughout the burst, rapid variability is observed across multiple energy bands, with the minimum variability timescale increasing from $\sim$2~ms during the early prompt phase to tens of milliseconds at later times \citep{2022_Troja, Veres_2023}.

Due to its proximity \citep[luminosity distance $d_L \simeq 350$~Mpc, $z=0.076$][]{Rastinejad_2022, 2022_Troja} and atypical long duration, GRB~211211A has one of the most detailed temporally resolved X-ray (15$-$150 keV, \swift-BAT) to $\gamma$-ray (10 keV - 10 MeV, \fermi-GBM) prompt emission spectra obtained to date. \citet{Gompertz2023} model the time-resolved spectra using a double smoothly broken power-law (2SBPL) model, in which each spectrum is described by power-law segments connecting two photon break energies, $E_{\rm p}$ and $E_{\rm b}$. They interpret these as two characteristic energies within a synchrotron framework. Specifically, $E_{\rm p}$ is associated with the characteristic synchrotron frequency of the minimum-energy electrons injected in the source, while $E_{\rm b}$ corresponds to the cooling frequency. The relative order of these frequencies determines whether the system is in the fast or slow cooling regime and their temporal evolution is found to be consistent with a transition of the system from fast to slow cooling. 

The goal of this work is to model the time-resolved prompt emission of GRB~211211A within a leptonic radiative framework, accounting for synchrotron radiation, inverse Compton scattering, and $\gamma \gamma$ pair creation. For this purpose, we compute numerical models that are fitted to time-average spectra in eight time intervals using  Bayesian inference. This approach allows us to infer the evolution of the physical properties of the prompt-emitting region during the first 60~s of the burst. Finally, we discuss the implications of the inferred parameters for the physical origin of the prompt emission and assess their consistency with expectations from relativistic magnetic reconnection.

This paper is structured as follows. 
We outline the main spectral and lightcurve characteristics of GRB~211211A in Section~\ref{sec:data}. We describe the model and fitting procedure in Section~\ref{sec:methods}. We present our results in Section~\ref{sec:results} and outline a physical framework for the prompt emission in Section~\ref{sec:minijet}. We discuss our results in Section~\ref{sec:discussion} and conclude in Section~\ref{sec:conclusions}. 

\section{Data} \label{sec:data}
We adopt the time-resolved spectral energy distributions (SEDs) presented in \cite{Gompertz2023}, which cover the first 60~s following the GBM trigger. We also include the broadband spectrum extracted during the 2$-$4~s interval. The time-resolved spectra were derived in eight time windows, with durations ranging from 2 to 20~s, using \textit{Swift} BAT (15$-$150~keV) and \textit{Fermi} GBM (10$-$900~keV) observations. 
Spectra were kindly provided by the authors.

For the purposes of this work, we assume that the entire 60~s emission belongs to the prompt phase. Although the origin of the late-time emission ($\gtrsim 12$~s) may differ from that of the main emission episode ($<12$~s) \citep[see][for an early-afterglow interpretation]{Veres_2023}, we find that the same radiative model provides an adequate description of the broadband spectra throughout the burst evolution (see Section~\ref{sec:results}).

\begin{figure}
    \centering
    \begin{overpic}[width=0.9\linewidth]{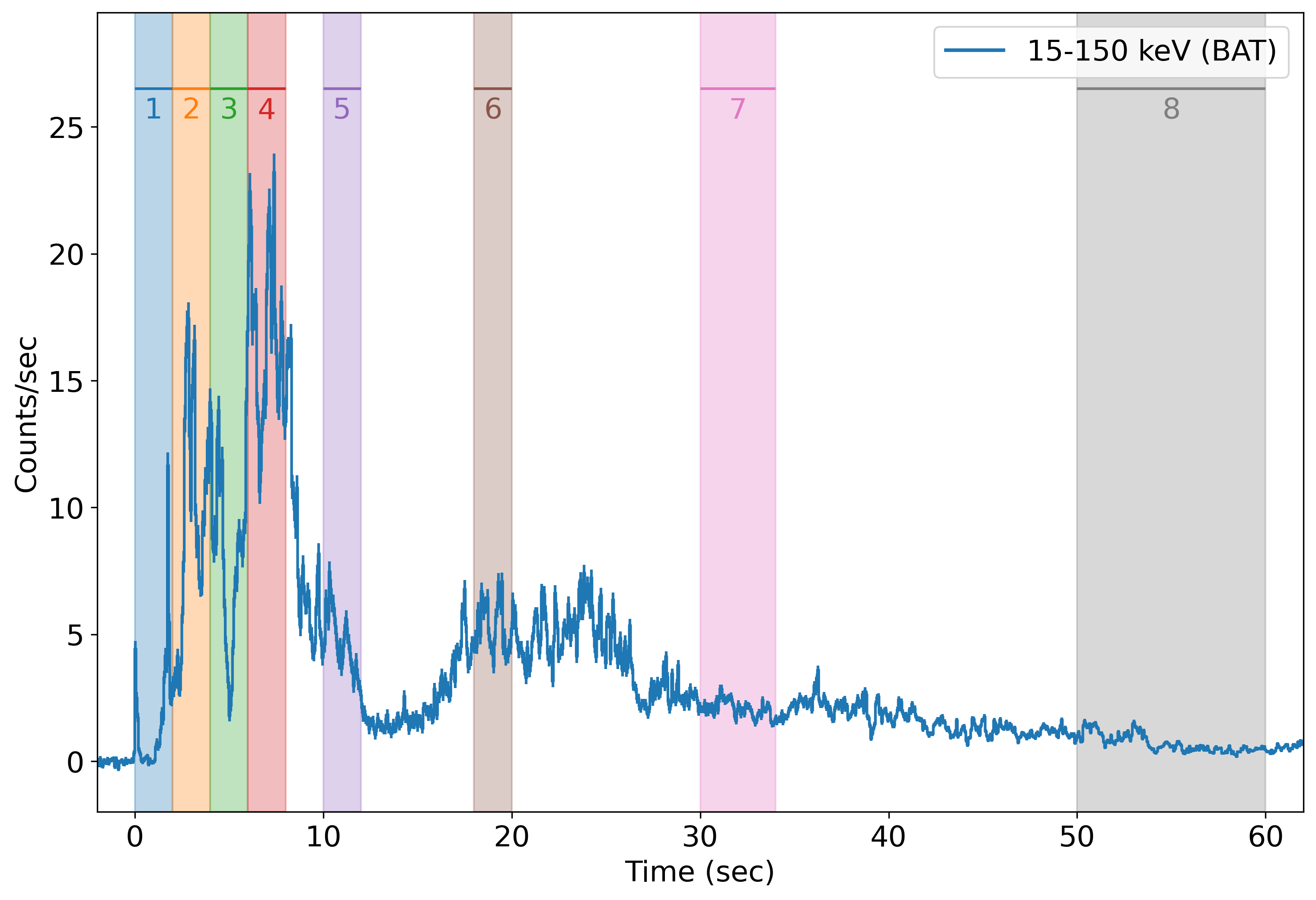}
        \put(59,35){%
            \begin{minipage}{0.25\linewidth}
                \centering
                \scriptsize
                \setlength{\tabcolsep}{4pt}
                \renewcommand{\arraystretch}{1.05}
                \begin{tabular}{cc}
                    \hline
                    Time interval & $\Delta t_{\rm obs}$ (ms) \\
                    \hline
                    $0-2$~s   & 2.6 \\
                    $2-4$~s   & 3.6 \\
                    $4-6$~s   & 6.9 \\
                    $6-8$~s   & 4.5 \\
                    $10-12$~s & 20 \\
                    $18-20$~s & 17 \\
                    $30-34$~s & 85 \\
                    $50-60$~s & 226 \\
                    \hline
                \end{tabular}
            \end{minipage}%
        }
    \end{overpic}
    \caption{64~ms binned \swift-BAT count rate light curve in the
    15$-$150 keV energy range. The vertical colored bands indicate
    the time intervals used in this work. The adopted variability
    timescales for each interval are also reported.}
    \label{fig:lightcurve}
\end{figure}

The flux exhibits rapid variability throughout the burst evolution. For example, this is evident in {\it Swift}-BAT light curves\footnote{\url{https://swift.gsfc.nasa.gov/results/batgrbcat/GRB211211A/data_product/01088940000-results/lc/}} across multiple energy channels and temporal binnings (1~s, 64~ms, 16~ms, and 8~ms). A dedicated analysis of the GBM light curves  
by \cite{Veres_2023} revealed a minimum variability timescale (MVT) of $\sim 2$~ms during the main emission episode (between 1 and 13~s after the trigger). Although the variability timescale was found to increase throughout the burst, it remained below 1 s.  We also note that the 10$-$900 keV (GBM) and 15$-$150 keV (BAT) light curves presented in \cite{Gompertz2023} were binned with a time resolution of 64~ms. In Fig.~\ref{fig:lightcurve} we present the 15$-$150 keV light curve, also binned with the same time resolution. The vertical colored bands correspond to the eight time intervals under study.

In our calculations, we adopt as the characteristic variability timescale $\Delta t_{\rm obs}$ the arithmetic mean of the MVTs reported by \cite{Veres_2023} for the individual time intervals (see their Figure~1). These MVTs range from $2.6$~ms for the $0-2$~s interval to $226$~ms for the $50-60$~s interval\footnote{The latter value corresponds to the MVT reported at $50$~s, as the analysis of \cite{Veres_2023} does not extend to later times.} -- see inset table in Fig.~\ref{fig:lightcurve}.

\section{Methods}\label{sec:methods}
\subsection{Model}\label{sec:model}
The GRB central engine launches a collimated outflow of magnetized plasma that is moving with Lorentz factor $\Gamma_{\rm j} \gg 1$ and has a half-opening angle $\theta_{\rm j}$.  The observed emission is assumed to originate within the relativistic outflow, where a fraction of the available energy is dissipated and transferred to relativistic particles. We model the emitting region as a homogeneous, relativistically moving spherical blob characterized by a small set of physical parameters.

In particular, the blob contains a tangled magnetic field of strength $B$ as measured in the blob rest frame. The blob radius is defined through the observed variability timescale $\Delta t_{\rm obs}$ as $R = \delta c \Delta t_{\rm obs}/(1+z)$.  

Here,  
the Doppler factor is defined as $\delta = \left[\Gamma\left(1 - \beta \cos{\theta_{\rm obs}}\right)\right]^{-1}$, where $\Gamma$ is the blob Lorentz factor\footnote{In general, the Lorentz factor of the blob can differ from the bulk Lorentz factor of the jet ($\Gamma \ne \Gamma_{\rm j}$). Our model probes the properties of the emitting blob.}, $\beta = \sqrt{1-1/\Gamma^2}$ and $\theta_{\rm obs}$ is the angle between the observer's line of sight and the jet axis. In the limit where $\Gamma \gg 1$ and $\theta_{\rm obs} \rightarrow 0$, the Doppler factor becomes $\delta \approx 2\Gamma$. 

Relativistic electrons are injected into the blob at a constant rate
with a power-law distribution in Lorentz factors,
\begin{equation}
    Q_{\rm e}(\gamma)= \frac{{\rm d}N_{\rm e}}{{\rm d}t {\rm d}\gamma} = Q_{\rm 0}\gamma^{-p}, \qquad \gamma_{\rm min} \leq \gamma \leq \gamma_{\rm max},
\end{equation}
where $\gamma_{\min}$ and $\gamma_{\max}$ are the minimum and maximum Lorentz factors of the distribution, $p$ is the power-law index, and $Q_{0}$ is the normalization that is defined by the total injected power as $L_{\rm e}=m_{\rm e}c^2\int_{\gamma_{\rm min}}^{\gamma_{\rm max}}\gamma Q_{\rm e}(\gamma){\rm d}\gamma$. The free parameters of the model are summarized in Table~\ref{tab:priors}. 

The leptonic radiative processes considered within the emitting region are synchrotron radiation, inverse Compton scattering, $\gamma\gamma$ pair production, and synchrotron self absorption. All particles may escape from the emitting region in a timescale $t_{\rm esc} \sim R/c$. 

\subsubsection{Physical interpretation and caveats of the model}
Our model remains agnostic to the specific dissipation scenario, such as  internal shocks \citep[e.g.,][]{1994_Rees_Meszaros, 1997ApJ...490...92K, 1998_Daigne, 2000_Spada, 2009A&A...498..677B} or magnetic reconnection \citep[e.g.,][]{Drenkhahn_2002, 2011_Zhang_Yan}. Therefore, the adopted geometry and dynamics provide an approximate description of the emitting region. 

In the internal-shock scenario, the radiation is produced in a relativistic moving shell. For an
on-axis observer, only the part of the shell within an angle $\sim1/\Gamma$ of the line of sight contributes to the observed emission; this region is causally connected and has a comoving transverse size $\sim r/\Gamma$, where $r$ is the shell radius. The blob can be regarded as a model of this
region for an on-axis
observer and a uniform jet with opening angle $\theta_{\rm j}> 1/\Gamma$ (see Appendix~\ref{app:blob_vs_shell} for a quantitative discussion). In the magnetic reconnection scenario, the emission may arise in plasmoids formed in the reconnection layer,
which are more naturally described as localized blobs. In this case,
$\delta$ should be interpreted as an effective Doppler factor that
includes the motion of the plasmoid relative to the jet (Section~\ref{sec:minijet}).

However, our model does not describe the dynamics of the emitting region. For instance, in the internal-shock picture, shell collisions and the arrival-time differences of photons emitted from different parts of the shell are not modeled; these effects shape the observed light curves and may affect the spectrum, particularly for structured jets. In the reconnection scenario, the growth, acceleration and mergers of plasmoids, as well as the
evolution and anisotropy of the particle distribution, are not included.

The parameters inferred from our time-average spectral fits (Section~\ref{sec:results}) should therefore be regarded as effective parameters describing the region that dominates the emission during the time interval considered, rather than properties of a series of colliding shells or plasmoids.

\subsection{Code}
We compute the numerical SEDs with an updated version of the time-dependent non-thermal radiation code \code \citep{LeHaMoC2024}\footnote{\url{https://github.com/Stamatis-Ilias/LeHaMoC-V2/tree/new-version}}. The code solves the coupled kinetic equations that govern the evolution of relativistic particle distributions (protons, electrons, positrons, photons, and neutrinos) in a homogeneous spherical emitting region. In this work, we use the leptonic configuration of the code, which includes the physical processes described in the previous section, augmented with some new features. 
The most relevant updates for the present calculations are the module for inverse Compton energy losses in the Klein-Nishina regime and the numerical acceleration of the calculations (see Appendix~\ref{app:code-updates}).

The code evaluates the number of photons per unit frequency $N_{\gamma}(\nu)\equiv {\rm d}N/{\rm d}\nu$, in the rest frame of the emitting region. We convert this to specific radiative luminosity $L_{\nu} = c h\nu  N_\gamma /R$. Then, the observed flux is calculated as $ \nu_{\rm obs} F_{\nu_{\rm obs}} = \delta^4\nu L_{\nu}/(4\pi d^2_{L})$ where $\nu_{\rm obs}=\delta\nu/(1+z)$.

\subsection{Fitting procedure}
We treat the emission within each time window independently and perform Markov Chain Monte Carlo (MCMC) fitting to the corresponding time-average SED using the same set of uniform priors for all epochs (see Table~\ref{tab:priors}). Since we are interested in modeling the average spectrum within each time window of duration $T_{w, \rm obs}$, we assume a constant injection power in relativistic particles throughout the interval. 

For a given set of model parameters, the system is evolved for $N$ light-crossing times, and the resulting SED is averaged over the respective time interval before being compared with the data. We define

\begin{equation}
N=\min\left(\frac{c T_{w,\rm obs}\delta}{R (1+z)},10\right)
=\min\left(\frac{T_{w,\rm obs}}{\Delta t_{\rm obs}},10\right),
\label{eq:timesteps}
\end{equation}
where the radius of the emitting region is defined through the variability timescale (see Section~\ref{sec:model}). We find that, depending on the model parameters, the coupled system of particles and photons reaches a quasi-steady state after approximately 3$-$5 light-crossing times. Beyond this point, the SED evolves only marginally, such that extending the calculation to later times has a negligible impact on the time-average spectrum. Since all observed time windows satisfy $T_{w,\rm obs}/\Delta t_{\rm obs}\gtrsim 30$, evolving the system for 10 light-crossing times provides a conservative estimate of the asymptotic spectral state while substantially reducing the computational cost of the MCMC calculations.

\begin{table}
    \caption{Uniform priors used in the MCMC fits.}
    \label{tab:priors}
    \centering
    \begin{tabular}{l l}
    \hline
     Parameter & Range  \\
     \hline 
     $\log_{10} \delta$    & $\mathcal{U}$[1.5, 3.6] \\ 
     $\log_{10} B$ (G)    & $\mathcal{U}$[-1.0, 3.0] \\ 
     $\log_{10} L_{\rm e}$~(erg/s) & $\mathcal{U}$[34.0, 42.0] \\ 
     $\log_{10} \gamma_{\min}$ & $\mathcal{U}$[2.0, 5.0] \\ 
     $\log_{10} \gamma_{\max}$ & $\mathcal{U}$[5.1, 7.0] \\   
     $p$ & $\mathcal{U}$[0.5, 5.0] \\ 
     $\ln f$ & $\mathcal{U}$[-6.0, 0.0] \\ 
     \hline
    \end{tabular}
\end{table}

Given a data set $D$ and a vector containing the model parameters $\vartheta = \{ \delta, B, \gamma_{\min}, \gamma_{\max}, p, L_{\rm e}, \ln f \}$, we define the likelihood function as

\begin{equation}
\ln{\mathcal{L}({\vartheta}|D)} =  -\frac{1}{2} \sum_{i}^{} \left[\frac{(\phi_{\rm m,i}-\phi_{\rm d, i})^2}{\sigma_{\rm tot, i}^2} + \ln{\sigma_{\rm tot, i}^2} \right], 
\end{equation}
where $i$ runs over the number of flux points, $\phi \equiv \log_{10}( \nu_{\rm obs} F_{\nu_{\rm obs}})$, the subscripts $m, d$ refer to the model expectation and data, respectively, and $\sigma_{\rm tot}$ is defined as

\begin{equation}
\label{eq:sigma_tot}
\sigma_{\rm tot, \rm i}^2 = \sigma_{\rm i}^2 + e^{2\ln{f}}.
\end{equation}

Here,  $\sigma_{\rm i}$ is the statistical uncertainty of the flux measurement $i$ (in logarithm). We also introduced the term $\ln f$ to account for other sources of uncertainty that are not captured by $\sigma_{\rm i}$, such as those arising from the combination of fluxes measured with different instruments.  

Upper limits on fluxes are also included in the fit by adding the following term to the likelihood function \citep[see][]{2012PASP..124.1208S,2020MNRAS.491..740Y}:
\begin{equation}
\sum_j^{} \ln \left ( \sqrt{\frac{\pi}{2}}\sigma_{\rm j} \left [1+\textrm{erf}\left(\frac{\phi_{\rm lim,j}-\phi_{\rm m,j}}{\sqrt{2} \sigma_{\rm j}}\right) \right] \right),
\end{equation}
where $\phi_{\rm lim, j}$ is the $j$-th flux upper limit (in logarithm) and $\sigma_{\rm j}$ is the respective uncertainty. 

We perform MCMC sampling using the {\tt emcee} algorithm \citep{2013PASP..125..306F}. We used 56 walkers that were propagated for 10,000 steps with the first 4000 steps being discarded as burn-in. The walkers were initialized by perturbing a common starting point according to $\theta_0 = \theta_{\rm init} + 10^{-4} \mathcal{N}(0,1)$, where $\theta_{\rm init}$ denotes the vector of initial parameters, and $\mathcal{N}(0,1)$ is a vector of random variables drawn from a normal distribution. The same initial parameter vector was adopted for all time intervals, while the individual walker positions were randomized independently.

We acknowledge that the present fitting procedure is not fully self-consistent and that a forward-folded analysis would be needed to assess its quantitative impact on the inferred parameters. A joint forward-folded analysis of \swift-BAT and \fermi-GBM data with a time-dependent SSC model would require a dedicated implementation of \code within a suitable fitting framework, such as GammaPy; this is still work in progress.

\section{Results} 
\label{sec:results}

We present the results of the fitting procedure described in the previous section, starting with the SEDs and continuing with the fitted model parameters.

\begin{figure*}
    \centering
    \includegraphics[width=0.9\textwidth]{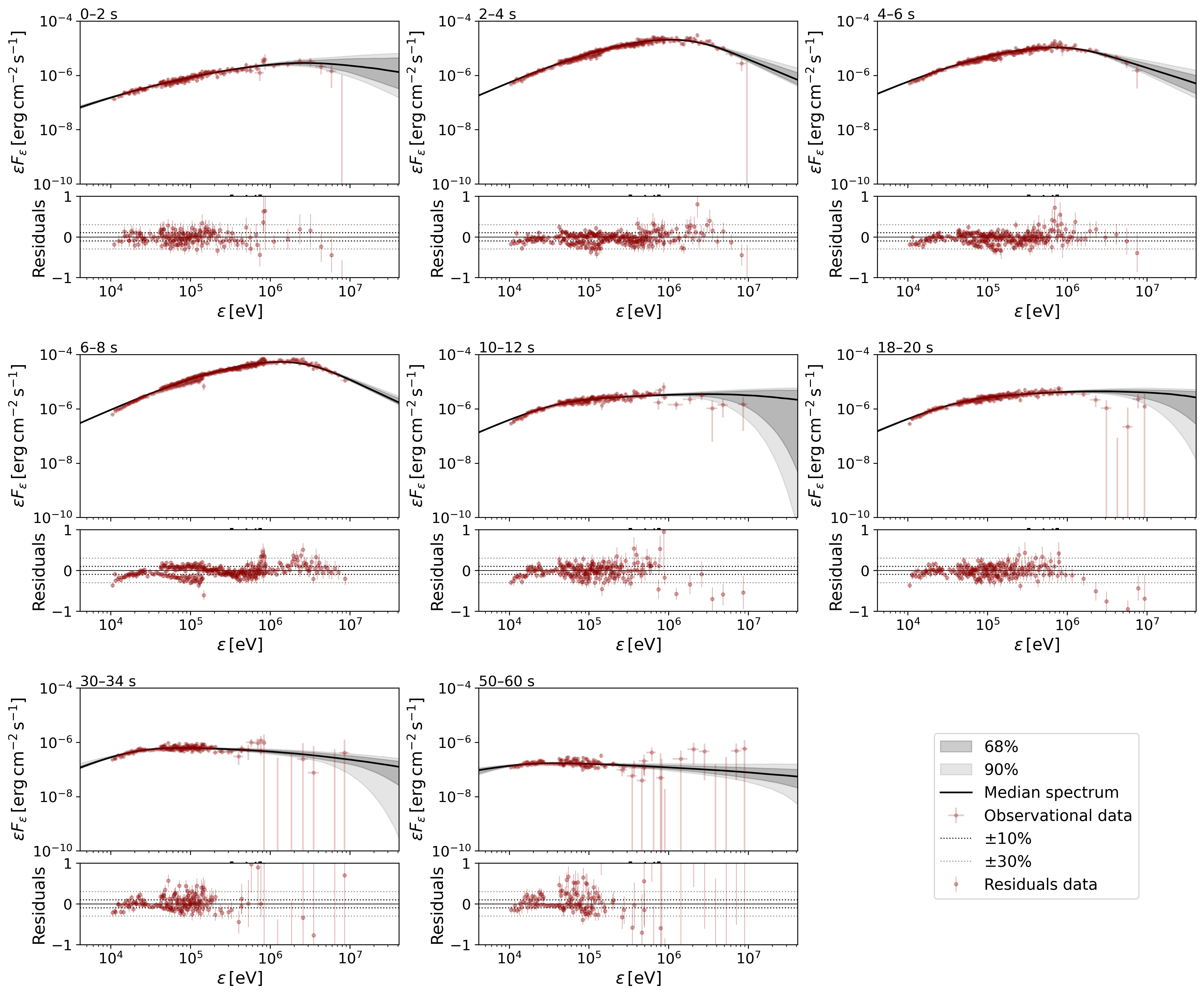}
    \caption{Observed and best-fitting numerical SEDs and residuals for the eight time intervals under study. In each panel, the observational data are shown as dark red markers, while the solid black line represents the median model, derived from 1000 randomly selected samples from the posterior distribution. The dark and light shaded regions represent the 68\% and 90\% credible intervals, respectively. The lower subpanels show the residuals between the observational data and the median model, as defined in \autoref{sec:results}. The horizontal lines indicate $\pm 10\%$ and $\pm 30\%$ relative deviations.}
    \label{fig:SEDs}
\end{figure*}

\begin{figure*}
    \centering
    \includegraphics[width=0.9\textwidth]{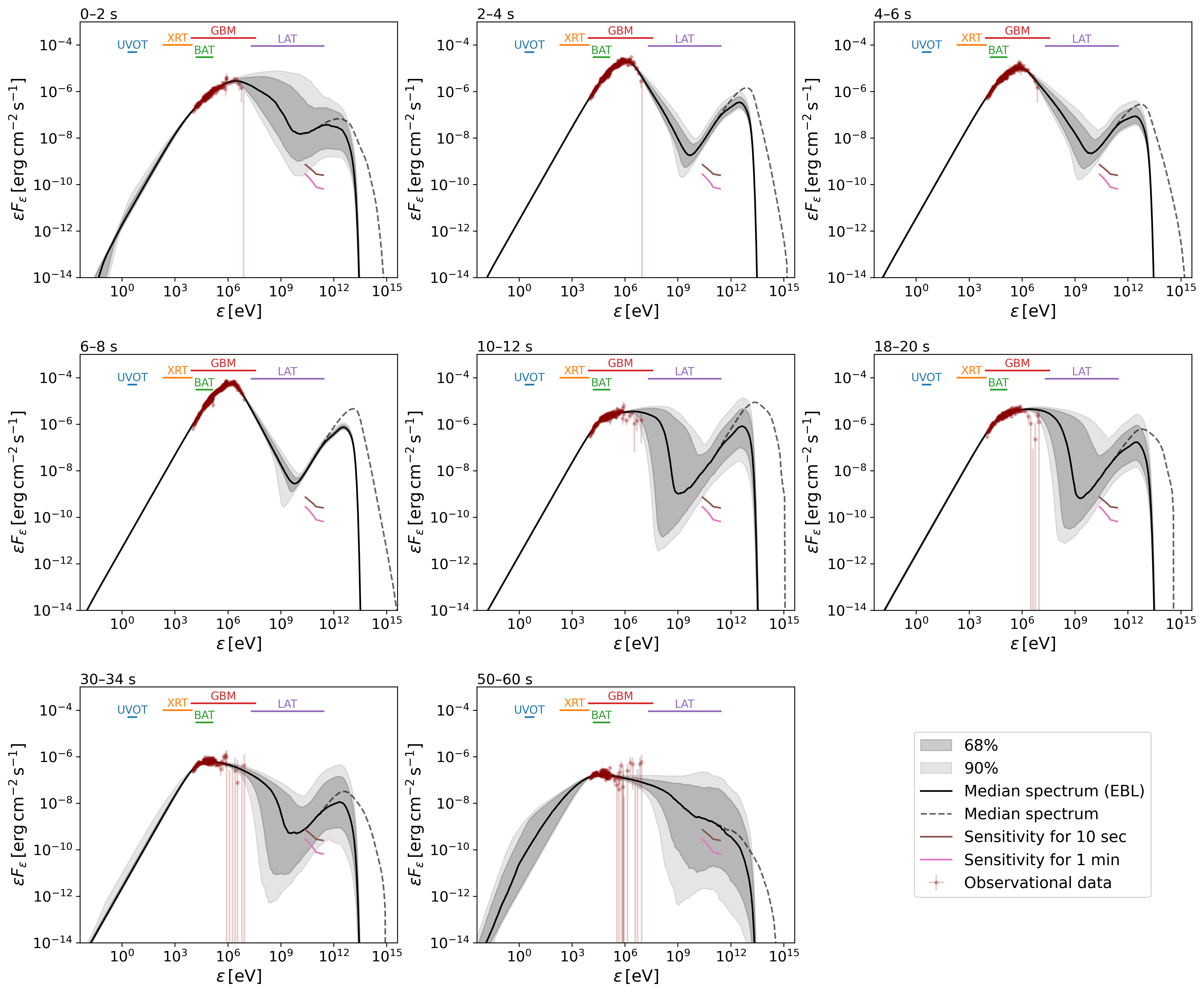}
    \caption{Observed and best-fitting numerical SEDs for the eight time intervals under study, taking into account EBL absorption effects. In each panel, the observational data are shown as dark red markers. The dashed gray line represents the median model derived from 1000 randomly selected samples from the posterior distribution without EBL attenuation, while the solid black line shows the corresponding median model after accounting for EBL absorption using the model of \cite{Finke_2022}. The dark and light shaded regions represent the 68\% and 90\% credible intervals, respectively. The brown and pink lines denote the CTAO sensitivity limits for the corresponding exposure times \citep{CTA_2019}. The colored bars indicate the frequency ranges covered by the instruments mentioned.}
    \label{fig:SEDs-full}
\end{figure*}

\subsection{Spectral energy distributions}
In Fig.~\ref{fig:SEDs} we present the observed and best-fitting numerical SEDs for all time intervals. The median model, together with the 68\% and 90\% credible regions, was derived from 1000 randomly selected samples drawn from the posterior distributions. To facilitate the comparison between the model and the data, we display the fractional residuals below each SED panel, which are defined with respect to the $ \nu_{\rm obs} F_{\nu_{\rm obs}}$ fluxes, namely 
$(10^{\phi_{d}}-10^{\phi_{\rm m}})/10^{\phi_{\rm m}}$. The residuals were computed using the median model interpolated at the frequencies of flux measurements. The residual values for low-frequency data remain largely within  $\pm 10\%$  throughout the first four time windows, corresponding to the early emission phase. The residuals in higher frequencies show a slight increase, although they remain within $\pm 30\%$. At later times ($>10$~s), a more systematic increase in the residuals becomes apparent, particularly for high-frequency observational data, indicating a discrepancy between the observations and the numerical SEDs. However, it is worth noting that these high-frequency measurements are associated with larger observational uncertainties.

The spectral peak energy $E_{\rm pk}$, i.e., the energy at which the $\varepsilon F_{\varepsilon}$ spectrum reaches its maximum, as well as the spectral break energy $E_{\rm b}$ identified by \cite{Gompertz2023} between $\sim20$ and 60~keV, are overplotted in the panels of Fig.~\ref{fig:SEDs} as a visual guide. If associated with the synchrotron cooling break, $E_{\rm b}$ could in principle provide an additional constraint on the model parameters and help reduce their degeneracy. This has been demonstrated by \cite{2019A&A...628A..59O} using a simplified synchrotron model, following the discovery of spectral breaks at keV-tens-of keV energies and their interpretation as the synchrotron cooling frequency \cite{2017ApJ...846..137O}.
In our model, however, the steady-state photon spectrum is obtained
self-consistently for a given set of parameters from the numerical
solution of the time-dependent equations for the electron and photon
equations, accounting for additional radiative processes, including inverse Compton scattering in the Thomson and Klein-Nishina regimes and $\gamma\gamma$ pair production. Therefore, we do not impose an interpretation of the observed spectral break prior to the fit, as doing so could bias the inferred posterior distributions. Instead, we assess the physical origin and inferred cooling-break energy of the best-fitting models a posteriori, as discussed in the following subsection and in Appendix~\ref{app:cool}.

In Fig.~\ref{fig:SEDs-full} we present the broadband SEDs including both synchrotron and synchrotron self-Compton (SSC) components. The SSC emission extends to very high energies (VHE, $E>100$~GeV). Although the model predicts emission in the Fermi LAT energy range (0.1$-$100~GeV) due to synchrotron and SSC radiation, there are no observational constraints available during the minute-long emission of this burst. The reason for the lack of LAT observations during the prompt phase is the large angular separation between the center of the LAT’s field of view  and the location of the burst at the time of the GBM trigger \citep[boresight angle of 106.5 degrees,][]{2021GCN.31210....1M}.

The extension of the SSC component to VHE makes the $\gamma$-ray attenuation due to extragalactic background light (EBL) an important effect. We account for this by adopting the EBL model of \cite{Finke_2022}. The impact of EBL attenuation is illustrated by overplotting the unattenuated median model with dashed lines. To assess the detectability of the predicted high-energy emission, we also show the differential flux sensitivity\footnote{\url{https://www.ctao.org/for-scientists/performance/}} of the Cherenkov Telescope Array Observatory (CTAO) North \citep{CTA_2019} at selected energies (25--250~GeV) for observation times of 10~s (brown lines) and 1~min (pink lines). For all time intervals considered, the predicted fluxes exceed the corresponding CTAO-North sensitivity limits, indicating that prompt VHE $\gamma$-ray emission from bursts with properties similar to those of GRB~211211A should be detectable with CTAO, provided that observations can start sufficiently early.  

\begin{figure}
    \centering
    \includegraphics[width=0.9\linewidth]{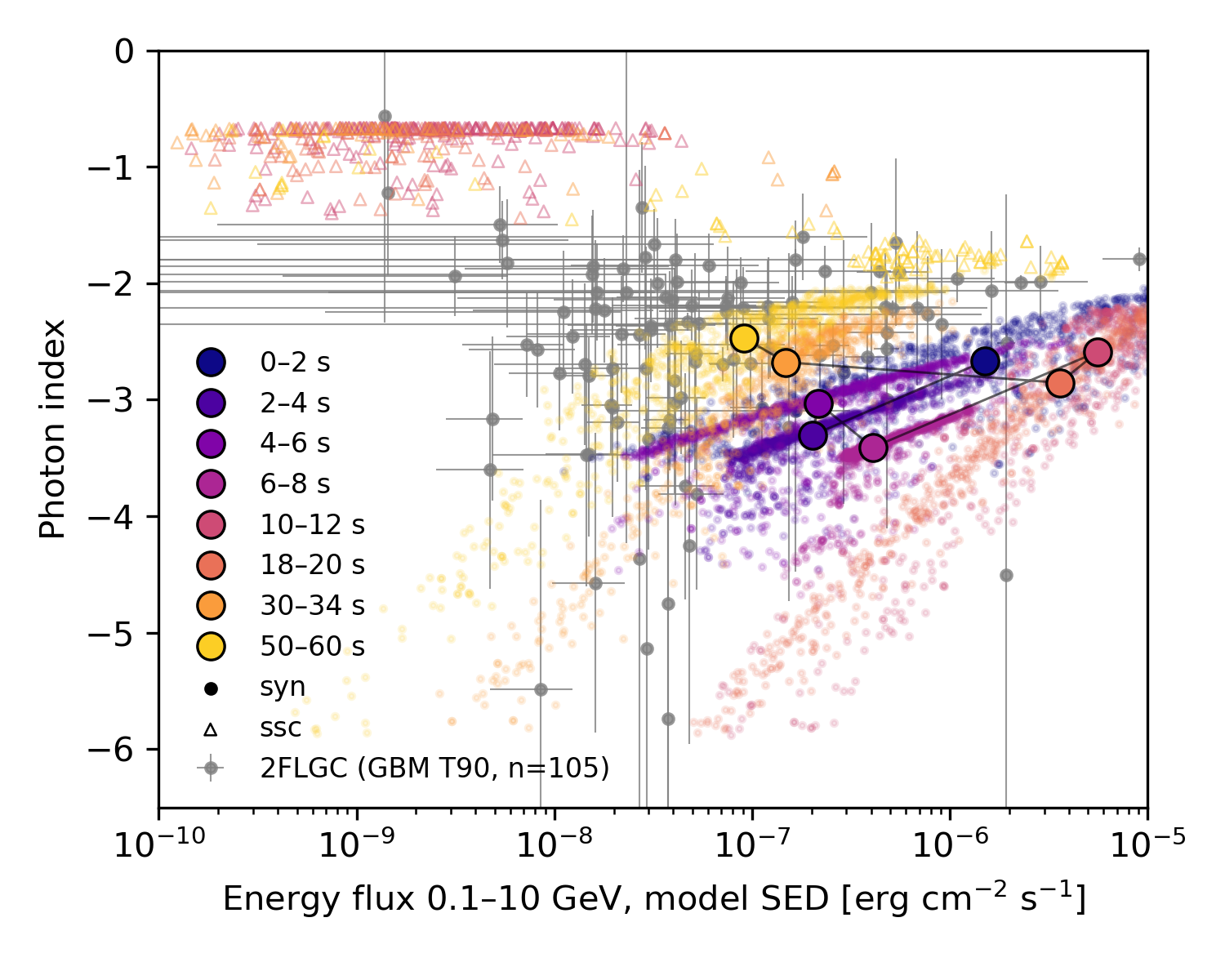}
    \caption{Photon index plotted against the integrated energy flux in the 0.1$-$10 GeV energy range. GRBs from the second Fermi-LAT GRB catalog \citep{2019ApJ...878...52A} that were detected by LAT within $T_{90}$ measured by Fermi-GBM are shown with grey markers. Colored points indicate our results from the analysis of SEDs from the posterior distributions of the fit. Each cloud contains about 1000 points, where filled circles indicate synchrotron-dominated solutions (with their mean values), while open triangles are the SSC-dominated solutions (see text for details).}
    \label{fig:LAT}    
\end{figure}

Despite the lack of LAT observations within the first minute of the burst, it is worth comparing the predictions of our model in the context of other LAT-detected GRBs. For this purpose, we selected bursts from the second Fermi-LAT GRB catalog \citep{2019ApJ...878...52A} with a reported photon index $>-5.9$ and with a LAT onset time occurring within the $T_{90}$ measured by GBM. This selection resulted in a subsample of 105 GRBs plotted with grey markers in Fig.~\ref{fig:LAT}. The colored points show our predictions for GRB~211211A. Specifically, we used 1000 SEDs drawn from the posterior distribution of each time-resolved fit. For each SED we computed the expected photon spectrum in 0.1–10 GeV and fitted it with a power law by maximising the Poisson likelihood of the model’s expected photon distribution. This yields the photon index that a LAT analysis would converge to in the high photon count limit, neglecting effects of the instrument response function. Energy fluxes were obtained by integrating the  model SED over the same band.

Each SED was further classified using the location of the $\varepsilon F_{\varepsilon}$ minimum, which marks the transition from the synchrotron tail to the rising SSC component. Spectra in which $\varepsilon F_{\varepsilon}$ decreases across the whole band are synchrotron-dominated, while those in which it monotonically increases are SSC-dominated. When the minimum lies within the band, the solution is SSC-dominated if more than half of the model’s photons lie above it, and a power law is fitted over the full energy range. Otherwise, a power law is fitted over the full range and below the minimum; if the two indices differ by less than 0.2, the SSC contribution does not affect the measured index and the solution is classified as synchrotron-dominated, with the index taken from the fit below the minimum. A small fraction of solutions ($<10\%$ in each interval) are cases in which a single power law does not provide a meaningful description, and they are not included in Fig.~\ref{fig:LAT}. 

The majority of solutions are synchrotron-dominated in the 0.1-10 GeV range. Most SSC-dominated solutions have hard indices ($\sim -0.7$), as expected for the rising low-energy side of the SSC component ($\varepsilon F_{\varepsilon} \propto \varepsilon^{4/3}$), and fluxes of $\sim 10^{-10}-10^{-8}~\rm erg \, cm^{-2} \, s^{-1}$. These are several orders of magnitude below the synchrotron emission of the early intervals and below typical LAT prompt-phase detections. We also note that the GeV emission detected several kiloseconds after the burst, having a flux of $10^{-9}~\rm erg~cm^{-2}~s^{-1}$  and a photon index of approximately $-3$, occupies a substantially different region of this parameter space when compared with our prompt-emission predictions \citep[see Extended Data Figure 2 of][]{Mei_2022}. Therefore, a prompt LAT detection would most likely probe the synchrotron spectrum, while the SSC component would be accessible to CTAO.

\subsection{Model parameters}
\begin{figure*}
    \centering
    \includegraphics[width=0.9\linewidth]{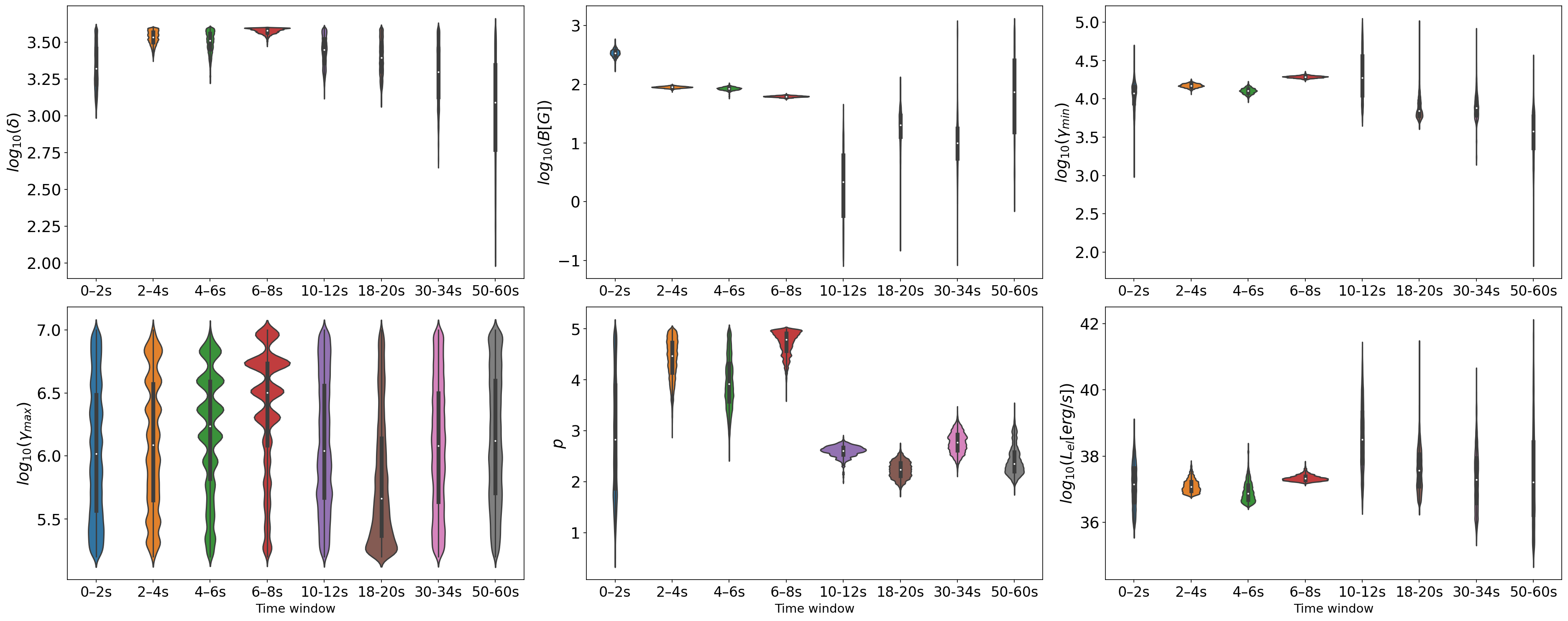}
    \caption{Violin plots showing the posterior distributions of the fitted model parameters during the burst evolution. White markers show the median values while thick (thin) bars indicate the 68\% (90\%) confidence intervals.}
    \label{fig:violin}
\end{figure*}

The fitted model parameters are summarized in Table~\ref{tab:results_MCMC}. Figure~\ref{fig:violin} presents the posterior distributions of the model parameters in the form of violin plots, allowing us to track their evolution throughout the burst. A key result of the SED modeling is the consistently large Doppler factors inferred for the emitting region, with median values ranging from $\sim 1400$ to $\sim 2500$ across all time intervals. The origin of these large Doppler factors can be understood using simple analytical arguments. Efficient synchrotron emission requires that the radiating particles -- including those injected with the minimum Lorentz factor -- cool before escaping the emitting region, while synchrotron losses remain the dominant cooling mechanism over SSC losses.

\begin{table*}
    \caption{Parameter table for the eight time windows. Reported values correspond to the median of the posterior distributions, while the quoted uncertainties represent the lower and upper bounds of the 68\% credible level.}
    \label{tab:results_MCMC}
    \centering
    \begin{tabular}{lllllllll}
    \hline
     \diagbox[width=4.0cm]{Parameter}{Time interval} & 0$-$2~s & 2$-$4~s & 4$-$6~s & 6$-$8~s & 10$-$12~s & 18$-$20~s & 30$-$34~s & 50$-$60~s \\
     \hline
     $\log_{10} \delta$ & $3.3^{0.2}_{0.2}$ & $3.53^{0.04}_{0.05}$ & 
     $3.51^{0.07}_{0.09}$ & $3.58^{0.01}_{0.02}$ & $3.5^{0.1}_{0.1}$ & $3.4^{0.1}_{0.1}$ & $3.3^{0.2}_{0.2}$ & $3.1^{0.4}_{0.5}$ \\[5pt]
     $\log_{10} B$ (G) & $2.53^{0.06}_{0.07}$ & $1.95^{0.01}_{0.02}$ & 
     $1.93^{0.02}_{0.03}$ & $1.79^{0.01}_{0.01}$ & $0.2^{0.7}_{0.8}$ & $1.3^{0.2}_{0.4}$ & $1.0^{0.4}_{0.4}$ & $1.8^{0.8}_{0.9}$ \\[5pt]
     $\log_{10} L_e$~(erg/s) & $37.2^{0.7}_{0.8}$ & $37.1^{0.2}_{0.2}$ & $36.9^{0.4}_{0.3}$ & $37.32^{0.11}_{0.07}$ & $39^{1}_{1}$ & $37.6^{0.7}_{0.6}$ & $37.4^{0.9}_{0.9}$ & $37^{2}_{1}$ \\[5pt] 
     $\log_{10} \gamma_{\min}$ & $4.1^{0.1}_{0.3}$ & $4.18^{0.02}_{0.03}$ & $4.10^{0.04}_{0.04}$ & $4.29^{0.01}_{0.02}$ & $4.3^{0.4}_{0.4}$ & $3.9^{0.2}_{0.1}$ & $3.9^{0.2}_{0.1}$ & $3.6^{0.3}_{0.3}$ \\[5pt] 
     $\log_{10} \gamma_{\max}$ & $6.0^{0.7}_{0.6}$ & $6.1^{0.6}_{0.6}$ & $6.3^{0.4}_{0.6}$ & $6.5^{0.3}_{0.8}$ & $6.1^{0.7}_{0.5}$ & $5.6^{0.8}_{0.4}$ & $6.1^{0.6}_{0.6}$ & $6.1^{0.7}_{0.5}$ \\[5pt]   
     $p$ & $3^{2}_{1}$ & $4.5^{0.4}_{0.5}$ & $3.9^{0.7}_{0.5}$ & $4.8^{0.1}_{0.3}$ & $2.61^{0.09}_{0.11}$ & $2.3^{0.2}_{0.2}$ & $2.8^{0.2}_{0.2}$ & $2.4^{0.4}_{0.2}$ \\ [3pt]
     \hline
     \multicolumn{9}{l}{\textit{Derived quantity}} \\[3pt]
     $\log_{10} R$~(cm) &  $11.2^{0.2}_{0.2}$ & $11.53^{0.04}_{0.05}$ & $11.79^{0.07}_{0.09}$ & $11.68^{0.01}_{0.02}$ & $12.19^{0.11}_{0.12}$ & $12.1^{0.1}_{0.1}$ & $12.7^{0.2}_{0.2}$ & $12.9^{0.4}_{0.5}$ \\ [3pt]
     \hline
    \end{tabular}
\end{table*}

For solutions to be synchrotron dominated, the synchrotron photon energy density must not exceed the magnetic energy density, i.e., $u_{\rm syn} \le u_{\rm B} = B^2/(8\pi)$, where $u_{\rm syn} \propto (L_{\rm pk,obs})/(\delta^6 \Delta t^2_{\rm obs})$. This condition places a lower limit on the Doppler factor, $\delta$, for any given magnetic field strength, $B$. An additional constraint follows from the fast-cooling requirement. In particular, electrons with Lorentz factor $\gamma_{\rm cool}$, for which the synchrotron cooling timescale equals the escape timescale, emit synchrotron photons with characteristic energy $E_{\rm b}= \delta \gamma^2_{\rm cool} (B/B_{\rm cr}) m_e c^2$, where $B_{\rm cr} \equiv 2 \pi m_e^2 c^3/(he)$.  Combining these two requirements yields the following lower limit on the Doppler factor,

\begin{equation}
    \delta \gtrsim 3000 \left(\frac{L_{\rm pk,obs}}{10^{51}\, \rm erg \, s^{-1}}\right)^{3/16} \left(\frac{E_{\rm b}}{65\, \rm keV}\right)^{1/8} \left(\frac{\Delta t_{\rm obs}}{4.5\, \rm ms}\right)^{-1/8}
    \label{eq:doppler_factor}
\end{equation}

where $L_{\rm pk,obs}$ is estimated by the maximum  luminosity measured from the observational data. The reference values of $L_{\rm pk,obs}$, $E_{\rm b}$, and $\Delta t_{\rm obs}$ correspond to the $6-8$~s time interval.

The blob radius, which is an inferred quantity, remains relatively small during the first 8~s of the burst, ranging from $\sim 1.5 \cdot 10^{11}$~cm to $\sim 6 \cdot 10^{11}$~cm. As time progresses, the radius increases by approximately an order of magnitude, reaching its maximum value of $\sim 8 \cdot 10^{12}$~cm in the final time interval. This evolution is primarily driven by the dependence of $R$ on the variability timescale, which exhibits an increasing trend throughout the analyzed time intervals, as shown in Figure~\ref{fig:lightcurve}. Physically, a larger emitting region corresponds to longer electron and photon escape timescales, which can favor slow cooling solutions. However, the cooling regime also depends on other model parameters, which must be considered to determine whether the electron population is in the fast or slow cooling regime. This is discussed in detail in Appendix~\ref{app:cool}.

The minimum Lorentz factor of the injected electron distribution is constrained to values well above unity and remains approximately constant, within the 68\% credible intervals, up to $\sim$6$-$8~s after the trigger. However, at later times, it decreases by a factor of $\sim 3$. This evolution mirrors the temporal behavior of the spectral peak energy reported by \cite{Gompertz2023} and presented in their Figure~3. In contrast, the maximum Lorentz factor is poorly constrained, which is reflected in the wide spread of model predictions at energies beyond the spectral peak (see Fig.~\ref{fig:SEDs-full}).

The power-law slope of the injected particle distribution is also weakly constrained during the early phases of the burst. This is expected because the electron power-law slope is primarily determined by the spectral slope above the peak energy of the photon spectrum, where the available data span a relatively narrow dynamic range. At later times (starting after 10~s), the photon spectra exhibit a more extended and well-defined power-law segment, resulting in significantly tighter constraints on the particle index ($p\sim$2$-$3). 

Finally, the magnetic field strength varies throughout the burst. During the first few time intervals (up to 6$-$8~s), it remains relatively constant and is constrained within a comparatively narrow range ($\sim$100$-$300~G). At later times, however, the constraints become progressively weaker. Nevertheless, there is an increasing trend of the median value of the magnetic field strength, ranging from $\sim 1.6$~G during 10$-$12~s to $\sim 60$~G at the last time interval.
We also note that the temporal evolution of the magnetic field appears to be anti-correlated with that of the particle injection power, with intervals characterized by stronger magnetic fields generally requiring lower injection powers. This behavior is qualitatively expected, since particles of a given Lorentz factor radiate more efficiently via synchrotron emission in regions of enhanced magnetic field strength.

\begin{figure}
    \centering
    \includegraphics[width=0.80\linewidth]{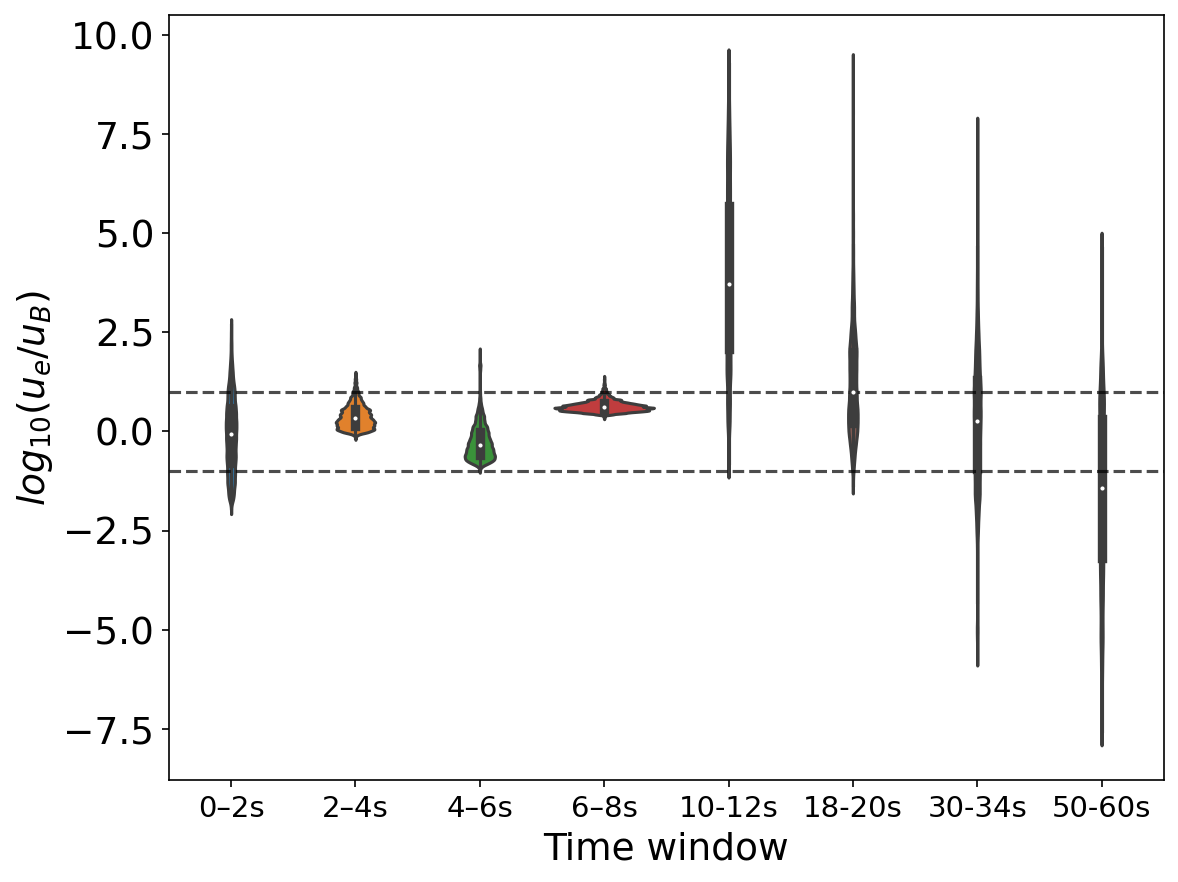}
    \caption{Violin plots showing the posterior distribution of the energy density ratio between relativistic pairs and the magnetic field in the emitting region across the evolution of the burst. White markers show the median values while thick (thin) bars indicate the 68\% (90\%) confidence intervals. Horizontal lines mark the range of values between 0.1 and 10.}
    \label{fig:violin-densities}
\end{figure}

In our radiative model, the magnetic field strength, $B$, and the power in relativistic electrons, $L_{\rm e}$, are independent parameters that are inferred from the MCMC fit. Consequently, the model does not make any prior assumption regarding the partition of energy between magnetic fields and relativistic particles. We define the particle energy density in terms of $L_{\rm e}$ as $u_{\rm e}=3L_{\rm e}/(4\pi R^2c)$, which represents the energy density that is available for radiation. In Fig.~\ref{fig:violin-densities} we present the posterior distribution of the particle-to-magnetic energy density ratio throughout the evolution of the burst. 
With the exception of two time intervals ($10-12$~s and $18-20$~s), the median value of the ratio lies between 0.1 and 10, suggesting rough energy equipartition between particles and the magnetic field in the emitting region. 
During the two intervals where $u_e/u_B > 10$, the corresponding SSC luminosity remains comparable to the synchrotron luminosity (see Fig.~\ref{fig:SEDs-full}). This suggests that inverse Compton scatterings that contribute to the peak SSC emission are occurring in the Klein-Nishina regime, which suppresses the Compton radiative efficiency despite the large particle energy density. 

\begin{figure}
    \centering
    \includegraphics[width=0.9\linewidth]{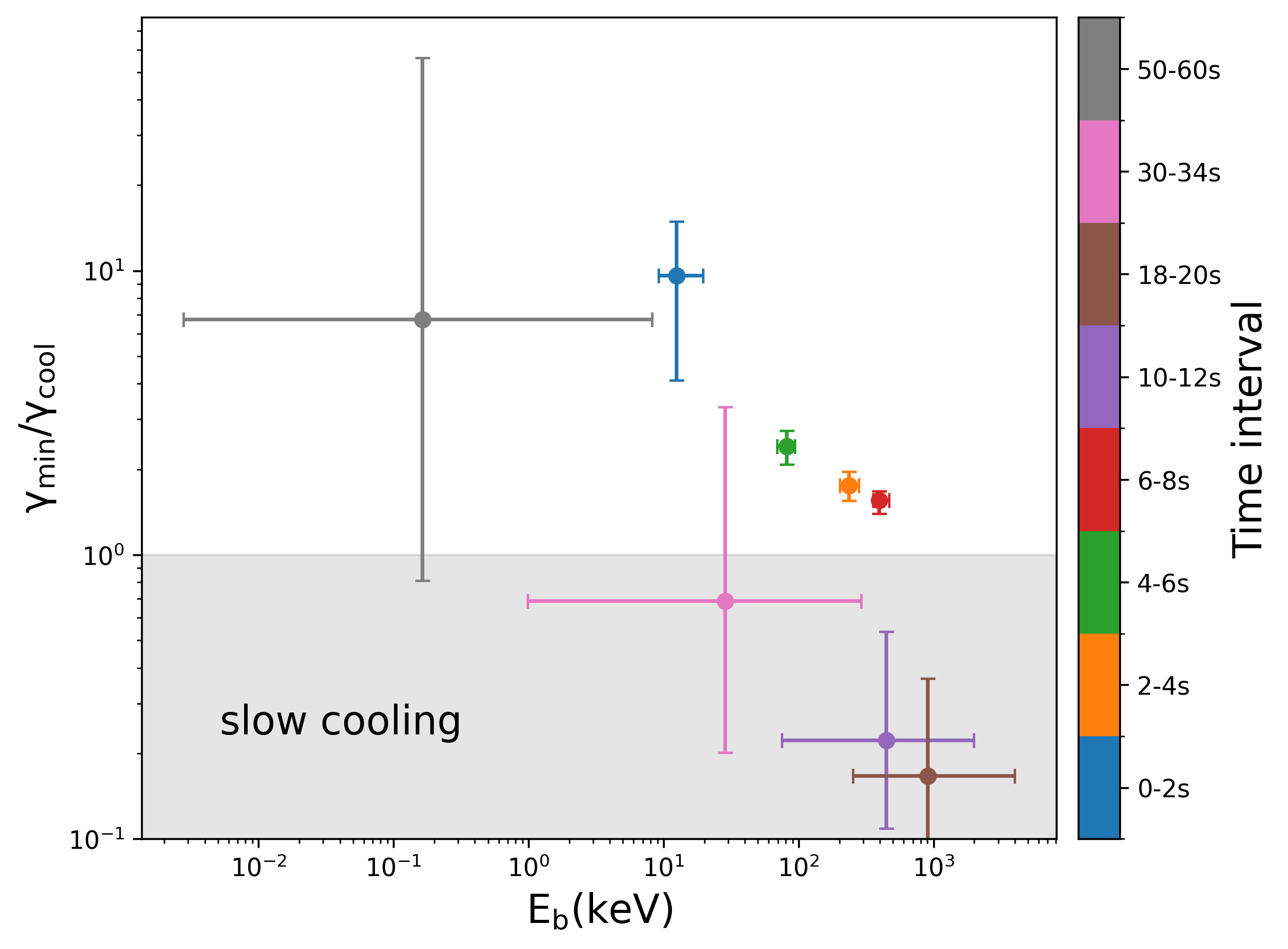}
    \caption{$\gamma_{\rm min}/\gamma_{\rm cool}$ ratio plotted against the cooling energy, $E_{\rm b}$. The points are color-coded according to each of the eight corresponding time intervals under study. The shadowed area, which covers ratio values less than 1, corresponds to a slow-cooling state of the system.}
    \label{fig:cool}    
\end{figure}

Having derived the posterior distributions for the fitted model parameters, we can evaluate the synchrotron cooling energy, $E_{\rm b}= \delta \gamma^2_{\rm cool} (B/B_{\rm cr}) m_e c^2$, where $\gamma_{\rm cool}$ is defined in Appendix~\ref{app:cool}. In Fig.~\ref{fig:cool} we plot the ratio of $\gamma_{\min}/\gamma_{\rm cool}$ against $E_{\rm b}$, with the error bars indicating the 68\% range of model predictions. The results are color-coded according to the corresponding time interval. Our modeling indicates that the radiating particles are in the fast-cooling regime during the first 8 seconds of the burst, evolving towards a marginally fast cooling regime ($\gamma_{\min}/\gamma_{\rm cool} \sim 1$). At later times, with the exception of the final time interval, which is also the least constrained, the particle population is instead in the slow-cooling regime. The transition from fast to slow cooling is qualitatively consistent with the model-independent spectral-fitting results of \cite{Gompertz2023}, although the specific time intervals associated with the two cooling regimes differ between the two analyses.Taken together, the posterior distributions of the fitted parameters and the evolution of $\gamma_{\min}/\gamma_{\rm cool}$ suggest that the burst can be broadly divided into two distinct phases: an initial phase lasting up to $\sim$ 8~s, encompassing the bright pulses of the light curve, followed by a second phase characterized by lower-amplitude pulses extending to $\sim 60$~s.

It is noteworthy that the values of $E_{\rm b}$ during the fast-cooling phase of the burst lie above 10~keV, consistent with the highest values obtained by \cite{2019A&A...628A..59O} in their analysis of a sample of 34 GRBs observed by XRT during the prompt phase. The requirement of high bulk Lorentz factors to suppress excessive SSC cooling is another common feature of the two analyses, although the values inferred in our modeling are even more extreme, owing in part to the much shorter variability timescales inferred for GRB~211211A. Our inferred magnetic field strengths during the fast-cooling phase are also substantially higher than those reported by \cite{2019A&A...628A..59O}. This difference, however, does not indicate a tension between the two results, but is a natural consequence of the different variability timescales considered in the two analyses. In \cite{2019A&A...628A..59O}, the default variability timescale used was 1~s, whereas the much shorter variability timescale observed in GRB~211211A implies a more compact emitting region. For a given observed luminosity, this requires a higher synchrotron emissivity, and hence a stronger magnetic field, to reproduce the observed emission. Thus, the higher magnetic field strengths inferred for GRB~211211A are consistent with the shorter variability timescale of this burst and should not be interpreted as a discrepancy with the values reported by \cite{2019A&A...628A..59O}. Overall, the fast-cooling phase of the prompt emission in GRB~211211A exhibits properties similar to those found in GRBs showing evidence for synchrotron cooling breaks in their prompt X-ray spectra.

\section{Relativistic blobs within the jet}\label{sec:minijet}
A robust finding of the SED modeling presented in Section~\ref{sec:results} is the inference of very large Doppler factors for the emitting region, with median values ranging from 1400 up to 2500. These translate to jet Lorentz factors $\Gamma_{\rm j}\sim$ 700$-$1250 assuming an on-axis observer ($\theta_{\rm obs} \sim 0$~deg). Although such values are expected in GRB prompt emission synchrotron models \citep[e.g.,][]{Beniamini_2013, Burgess_2020}, such extreme Lorentz factors are difficult to reconcile with those inferred from afterglow observations of GRB~211211A. In particular, broadband modeling of the X-ray-to-radio counterpart at times $\gtrsim 0.1$~d, including the contribution from a kilonova component, favors an on-axis ($\theta_{\rm obs} = 0.688^{+0.401}_{-0.344}$) jet with Lorentz factor $73.11^{+51.70}_{-22.08}$~\citep{Rastinejad_2022}; see also \cite{Hamidani_2024} for similar constraints on the jet properties. Therefore, the Lorentz factors required by our prompt-emission modeling exceed those inferred from the afterglow by approximately an order of magnitude.

One possible way to alleviate this tension is to consider that the region responsible for the prompt emission is moving relativistically within the jet rest frame. Similar proposals have been made to explain fast flux variability from blazar jets \citep[e.g.][]{Giannios_2009, Nalewajko_2011, Narayan_Piran_2012, Giannios_2013, Petropoulou_2016, Christie_2019}, but have not been widely discussed in the context of GRB prompt emission \citep[e.g.][]{Lazar_2009, Barniol_Duran_2016, Burgess_2020}. In this section, we introduce a physical framework presented in \cite{Petropoulou_2016} to interpret the results obtained from the SED modeling of the prompt emission. 

The GRB jet is assumed to be a collimated, relativistic, strongly magnetized outflow with bulk Lorentz factor $\Gamma_{\rm j} \gg 1$. 
We assume that the prompt emission is powered by relativistic particles contained in plasma blobs (plasmoids) that form in reconnection layers within the GRB jet where the plasma magnetization is still large ($\sigma \gg 1$). Plasmoids move at an angle $\theta_{\rm co}$ with respect to the jet axis as measured in the rest frame of the jet. 

Particle-in-Cell (PIC) simulations of relativistic magnetic reconnection have shown that plasmoids are accelerated along the reconnection layer to velocities approaching the Alfv{\'e}n velocity, which itself is approaching the speed of light in high-$\sigma$ plasmas. The asymptotic Lorentz factor of the plasmoids (as measured in the rest frame of the jet) is $\Gamma_{\rm co} \approx \sqrt{\sigma} \gg 1$\footnote{This scaling applies for small and fastest plasmoids in the limit where the guide field (i.e., the magnetic field component that does not reconnect) is weak with respect to the reconnecting magnetic field.} \citep[e.g.][]{Sironi_2016}. The relativistic motion of plasmoids within the jet naturally enhances the effective Doppler boosting of their emission, as we demonstrate below. Therefore, it provides a plausible mechanism for reconciling the large Doppler factors inferred from prompt-emission modeling with the substantially lower bulk Lorentz factors derived from afterglow observations.

Let the jet axis define the $z$-direction of a Cartesian coordinate system. Without loss of generality, we assume that the plasmoid motion is confined to the $y-z$ plane. We denote the plasmoid velocity (in units of $c$) in the observer’s frame by $\beta_p$, with components parallel and normal to the jet axis given by
\begin{eqnarray}
    \beta_{p,\parallel} &=& \frac{\beta_j + \beta_{\rm co} \cos \theta_{\rm co}}{1 + \beta_j \beta_{\rm co}\cos \theta_{\rm co}} \\ 
    \beta_{p, \perp} &=& \frac{\beta_{\rm co} \sin \theta_{\rm co}}{\Gamma_j (1 + \beta_j \beta_{\rm co}\cos \theta_{\rm co})},
\end{eqnarray}
where $\beta_j\sim 1$ is the speed of the jet plasma (in units of $c$). The Lorentz factor of the plasmoid, as measured in the observer's frame, is then written as:
\begin{equation}
    \Gamma_p = \Gamma_j \Gamma_{\rm co} (1 + \beta_j \beta_{\rm co}\cos \theta_{\rm co}).
\end{equation}
Finally, the Doppler factor of the plasmoid, which is the parameter we are probing with SED fitting, is given by 
\begin{equation}
    \delta_p = \frac{1}{\Gamma_p (1 - \beta_p \cos \chi)},
\end{equation}
where $\cos \chi = \cos(\theta_{\rm obs} - \theta)$, $\theta_{\rm obs}$ is the observer's viewing angle and $\theta$ is the angle between the jet axis and the plasmoid direction of motion in the observer's frame
\begin{equation}
    \tan \theta = \frac{\beta_{\rm co} \sin \theta_{\rm co}}{\Gamma_j (\beta_j + \beta_{\rm co} \cos \theta_{\rm co})}.
\end{equation}

\begin{figure}
\centering
\includegraphics[width = 0.9\linewidth]{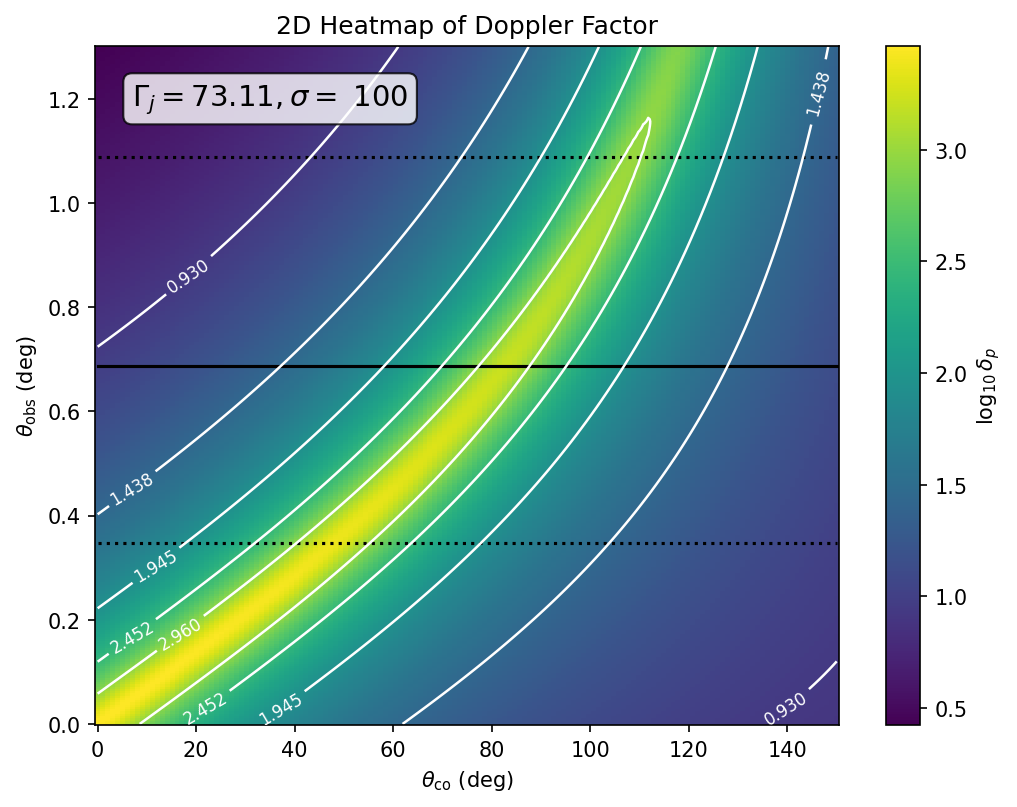}
\caption{Heatmap of the plasmoid Doppler factor expected for different directions of motions (measured by the angle $\theta_{\rm co}$ in the jet reference frame) $\theta_{\rm co}$ and observer's viewing angles $\theta_{\rm obs}$. The best-fit value for the viewing angle and the 68\% confidence interval, as inferred from afterglow modeling \citep{Rastinejad_2022}, are indicated with horizontal solid and dotted lines, respectively.}
\label{fig:heatmap}
\end{figure}

Using $\Gamma_j = 73.11$ as a reference value inferred from afterglow modeling \citep{Rastinejad_2022} and $\sigma=100$ as an indicative value for the jet magnetization, we present in Fig.~\ref{fig:heatmap} a heatmap of the Doppler factor of the plasmoid for different combinations of the two angles $\theta_{\rm co}$ and $\theta_{\rm obs}$. The figure demonstrates the strong dependence of the Doppler factor on both angles, suggesting that plasmoids moving at $\sim 80$~deg with respect to the jet axis (in the jet reference frame)\footnote{In the observer's frame the blobs are almost aligned to the observer's line of sight, namely  $\theta \approx 1/\Gamma_j \approx \theta_{\rm obs}$.} can result in the high Doppler factor values inferred from SED fitting for $\theta_{\rm obs}\sim 0.7$~deg.

In Appendix~\ref{app:minijet} we derive constraints on the jet plasma magnetization $\sigma$ and the direction of motion of plasma blobs in the jet, which are presented in Fig.~\ref{fig:corner-blob}.  

\section{Discussion} \label{sec:discussion}
In this work, we modeled the time-resolved prompt emission spectra of GRB~211211A by incorporating the leptonic radiative processes of synchrotron radiation, inverse Compton scattering, $\gamma\gamma$ pair production, and synchrotron self absorption. For each time interval during the prompt emission phase, we modeled the emitting region as a homogeneous, relativistically moving spherical blob associated with the GRB outflow. The corresponding spectral energy distributions were computed using \code, a time-dependent non-thermal radiation code.

We inferred the physical properties of the emitting regions (blobs) by modeling the time-average SEDs extracted over eight temporal intervals each lasting for several seconds (see Table~\ref{tab:results_MCMC}). The light curves exhibit pronounced variability, particularly during the main emission episode (approximately 2$-$12~s), indicating that the observed emission is likely produced by a substantially larger number of individual emitting regions. A more realistic description would therefore require time-resolved spectroscopy of individual pulses, allowing the physical properties of each emitting region to be inferred separately. In this sense, the present analysis provides the average properties of a population of emitting regions contributing to the emission within each time interval, rather than the properties of individual blobs. 
 
A related simplification of our modeling concerns the temporal evolution of the emitting region. For each MCMC realization, the emitting region is evolved for up to ten light-crossing times (see Eq.~\ref{eq:timesteps}) to compute the time-average spectrum, without accounting for adiabatic expansion. Incorporating adiabatic expansion of the emitting region over such an extended period would increase the source size and photon escape timescale substantially, leading to slower flux variations than observed in the data. This simplification, however, should not be interpreted as implying that a realistic blob survives for tens of dynamical times while maintaining a fixed size. A more realistic treatment would require modeling of spectra from individual pulses, where each pulse is associated with a blob that evolves for $\sim$~1$-$2 dynamical timescales. 

We find that the inferred very large Doppler factors may be generally reproduced by blobs (plasmoids) propagating at relatively large angles ($\sim 80-90$~deg) with respect to the jet axis in the jet comoving frame. This result suggests that the reconnection layers responsible for accelerating the plasmoids are oriented nearly perpendicular to the jet axis. Such a geometry is naturally expected if the magnetic field in the prompt-emission region is predominantly toroidal. Strong toroidal magnetic fields arise in several GRB jet scenarios, including striped-wind models \citep[e.g.,][]{Spruit_2001, Drenkhahn_2002, Giannios_2019} and MHD axisymmetric jets, where the toroidal component decays more slowly with distance than the poloidal field components \citep[e.g.,][]{Vlahakis_2003_1, Vlahakis_2003, Lyubarsky_2009, Bromberg_2016}.

The inferred plasma magnetization is $\sigma \sim 40-800$ with the highest values obtained between 2~s and 8~s and the lowest derived for 50$-$60~s (see Fig.~\ref{fig:corner-blob}). PIC simulations have shown that reconnection in such highly magnetized plasmas produces non-thermal pair distributions characterized by a break at $\gamma_{\rm br}\sim 0.1 \sigma_{\pm}$, where $\sigma_{\pm}$ is the pair plasma magnetization \citep[for a recent review see][]{Sironi_2025}. Assuming a cold proton--electron--positron plasma with lepton multiplicity $\kappa=(n_+ + n_-)/n_p$, the corresponding pair plasma magnetization is written as $\sigma_{\pm}=(\sigma/\kappa)(m_p/m_e)$, yielding $\sigma_{\pm}\sim10^4-10^5$ for $\kappa\sim 3-5$. 
Interestingly, the values of $\gamma_{\min}$ inferred from the SED fits (see Fig.~\ref{fig:violin}) are broadly consistent with the expected values of $\gamma_{\rm br}$ for $\kappa \sim 3$, once the uncertainties in $\sigma$ and $\gamma_{\min}$ are taken into account. 

We also note that the pair distributions predicted by kinetic simulations are generally described by broken power laws, whereas our modeling assumes a single power-law distribution above $\gamma_{\min}$. Since the distribution at energies below the break is typically very hard  ($p_< \sim 0.5-1$) for $\sigma_\pm \gg 1$, a significant fraction of synchrotron emission will arise from particles close to the break of the distribution. In this regard, the inferred values of $\gamma_{\min}$ may be viewed as a proxy for the characteristic break of the non-thermal pair distribution generated by relativistic magnetic reconnection. 

A common feature of relativistic magnetic reconnection is that the accelerated particles and magnetic fields within plasmoids are in rough energy equipartition, even when the upstream plasma is initially highly magnetized \citep[e.g.,][]{Sironi_2015, Petropoulou_2019}. Although a few time intervals favor particle-dominated solutions, the majority of the posterior distributions are broadly consistent with equipartition between relativistic pairs and magnetic fields (see Fig.~\ref{fig:violin-densities}). 

We note that our radiative transfer calculations assume an isotropic pitch-angle particle distribution. PIC simulations of relativistic reconnection  have demonstrated that particles can exhibit energy-dependent pitch-angle anisotropy \citep[e.g.,][]{Comisso_2023, Comisso_2024}. In particular, particles with Lorentz factors close to the break of the distribution tend to have smaller pitch angles, thus emitting less efficiently synchrotron radiation than in isotropic models. 
Given that particles with $\gamma_{\min}$ (which can be considered as the effective $\gamma_{\rm br}$ of the distribution) are found to contribute significantly at the spectral peak of synchrotron emission, reproducing the observed emission under anisotropic conditions would likely require stronger magnetic fields than those inferred here. This would, in turn, reduce the inferred values of $u_e/u_B$, bringing the emitting regions closer to energy equipartition.

Although the discussion above lends support to relativistic reconnection as a plausible physical interpretation of the prompt emission of this burst, a definitive assessment of this correspondence requires detailed modeling using broken power-law pitch-angle anisotropic distributions calibrated against PIC simulations.

The inferred posterior distributions may depend, to some degree, on the adopted priors and the initialization of the MCMC chains. As an illustrative example, imposing priors of $39 < \log_{10}(L_{\rm e} \rm [erg/s]) < 44$ and $-1 < \log_{10}(B [\rm {G}]) < 2$ led to a different family of solutions, which are slow-cooling and SSC-dominated.  Although the fits in the BAT and GBM energy ranges are statistically comparable to those presented in Fig.~\ref{fig:SEDs}, these models predict very bright SSC emission peaking at $\sim 100$~TeV and require extreme particle-to-magnetic energy densities, reaching values of $\sim 10^5-10^6$. While these solutions may be disfavored on physical grounds, their existence highlights the presence of multiple local minima in this high multi-dimensional parameter space. MCMC samplers, such as {\tt emcee}, may have difficulty transitioning between
widely disconnected local minima once the walkers have converged to a particular region of parameter space. 

In this work, we have used the efficiency of \code to fit multi-epoch SEDs with MCMC. Nevertheless, as discussed above, there is a degeneracy between our preferred solutions and those with SSC-dominated spectra. A more comprehensive exploration of the parameter space would require nested sampling algorithms, such as UltraNest \citep{Buchner2021}, which are better suited to identifying multiple disconnected modes in the posterior distribution. The substantially larger number of model evaluations required by such methods would likely require the use of precomputed model grids or surrogate models in place of direct \code calculations \citep{Tzavellas2024}.
Neither option is currently applicable, since, for example, existing neural network emulators are trained for the parameter space of blazar jets \citep{Tzavellas2024, Begue_2024, 2025arXiv251000126T}. Similarly, performing event-based model fitting rather than fitting unfolded data \citep[e.g.,][]{Rosillo_2025} could help account for the instrumental cross-calibration issues reported between BAT and GBM \citep[see][]{Gompertz2023}. This approach would be useful if we were fitting a limited energy band with an empirical model; in our case, however, parameter uncertainties are dominated by the lack of data between the GBM and LAT energy ranges (see Fig.~\ref{fig:SEDs-full}), and would therefore not alter our findings.

\section{Conclusions} \label{sec:conclusions}
The prompt emission of GRB~211211A between 10 keV and 10 MeV can be successfully reproduced by synchrotron radiation from a population of relativistic electrons. 
The accompanying synchrotron self-Compton emission extends to TeV energies, with predicted fluxes that would be detectable by CTAO for a burst similar to GRB~211211A, provided a sufficiently rapid response to a \fermi-GBM trigger or if the burst occurs within the CTAO field of view. Future searches of VHE emission from GRBs with a compact object merger origin, like GRB~211211A, will benefit from gravitational wave (GW) pre-merger alerts \citep[see][for a detailed analysis on detection prospects with CTAO and Einstein Telescope] {Banerjee_2023}.

The spectral evolution during the first minute of the burst reflects changes in the physical conditions of the emitting region. Our best-fit models favor fast-cooling solutions for the first 8~s, followed by a transition to slow-cooling solutions at later times. The observed short variability of this burst requires very high Doppler factors ($\sim$ 1000$-$2500) throughout the burst evolution. Such extreme Doppler factors are difficult to reconcile with the jet Lorentz factor inferred from afterglow modeling unless the prompt-emitting regions are themselves moving relativistically with respect to the jet plasma. 

\section*{Author Contributions}
M.P. conceived and designed the project. She led the study by performing the SED modeling, interpreting the results within a physical framework, and preparing the first draft  of the manuscript. 
M.G. contributed to the analysis and interpretation of the results, the preparation of figures and supporting calculations, and the writing of the manuscript.
K.X. contributed to the writing of the Introduction and performed independent calculations that supported the validation of the results. His bachelor's research project laid the foundation for this study.
S.I.S developed, implemented, and validated the new \code module for electron cooling in the Klein-Nishina regime used in the modeling. He also contributed to the interpretation of the results and to the revision and editing of the manuscript. G.V. has provided advice on data analysis, statistical modeling, and fitting of the spectra, while also contributing to editing of the manuscript. A.M. has performed preliminary calculations and provided insightful comments and constructive feedback on the manuscript. 

\begin{acknowledgements}
The authors thank the anonymous referee for a constructive and insightful review. We also thank Gor Oganesyan for his positive feedback on the manuscript and for pointing us to several useful references. Benjamin P. Gompertz for kindly providing the time-average spectra used in this work. 
\end{acknowledgements}

\bibliographystyle{aa} 
\bibliography{refs.bib}

\appendix 
\section{Equivalence of the blob and shell geometries}
\label{app:blob_vs_shell}
\providecommand{\tvar}{\Delta t_{\rm obs}}

The radiative modelling presented in this work assumes a single, spherical,
homogeneous emitting region (``blob'') of radius $R$, moving with bulk Lorentz
factor $\Gamma$ towards the observer. 
In the standard picture of GRB prompt emission, however, the radiation is
produced in a thin, relativistically expanding spherical shell (or a portion of
it, for a jet with opening angle $\theta_{\rm j}\gtrsim 1/\Gamma$).

Here we show that, for a given luminosity and variability timescale, the two
geometries yield comoving energy densities that agree to within a factor of
order unity  for an on-axis observed of a uniform jet. 
This justifies the use of the (simpler) blob geometry for
computing quantities that depend on the comoving energy density, such as the
$\gamma\gamma$ and synchrotron self-absorption opacities, the radiative cooling
rates and the target photon fields for inverse Compton scattering and
\citep[see also][Chapter 5.5]{Dermer2009}. Throughout, we
compare bolometric quantities, so that the results do not depend on the shape of the spectrum.

As in the main text, unprimed quantities refer to the rest frame of the emitting region (blob or shell) and quantities with the subscript ``obs'' to
the observer frame. The shell is described by its width $\Delta$, measured in
its rest frame, and by its radius $r$, measured in the frame of
the central engine; the isotropic-equivalent energy $E_{\gamma,\rm iso}$ and
bolometric luminosity $L_{\gamma,\rm iso}$ also refer to the latter frame,
with $L_{\gamma,\rm iso} = 4\pi d_L^2\,\Phi_{\rm obs}$. All energy densities
below refer to the $\gamma$-ray photons in the rest frame of the emitting
region and are expressed in terms of the radiated (observed) $\gamma$-ray
luminosity.

First we consider a blob whose radius (in its rest frame) is set by the causality requirement
\citep[Eq.~5.44 of][]{Dermer2009}
\begin{equation}
R = \frac{\delta \, c\,\tvar}{1+z},
\label{eq:R_blob}
\end{equation}
where $\tvar$ is the observed variability timescale and
$\delta\simeq 2\Gamma$ for a blob moving towards the observer.
For isotropic emission in the blob frame, the $\gamma$-ray luminosity in the
blob frame is $L_\gamma = L_{\gamma,\rm iso}/\delta^4$
\citep[Eq.~5.46 of][]{Dermer2009}. Taking the photon residence time to be the
light-crossing time, $R/c$, the energy density in the blob
frame is
\begin{equation}
u_{\rm b} = \frac{L_\gamma}{V_{\rm b}}\,\frac{R}{c}
= \frac{3\, L_{\gamma,\rm iso}}{4\pi c\, R^{2}\, \delta^4}
\simeq \frac{3\,(1+z)^2\, L_{\gamma,\rm iso}}{256\pi\,\Gamma^6 c^3\,\tvar^2},
\label{eq:u_blob}
\end{equation}
where $V_{\rm b}=4\pi R^3/3$. With $L_{\gamma,\rm iso} = 4\pi d_L^2\,\Phi_{\rm obs}$,
Eq.~\ref{eq:u_blob} is equivalent to Eq.~5.48 of \citet{Dermer2009}.

Next we  consider a shell of width $\Delta$ (in its rest frame) located at radius $r$ (in the central-engine frame). Two timescales determine the duration of the pulse it produces: the radial one, set by the shell width, and the angular
(curvature) one, set by the arrival-time delay of emission from the visible
region $\theta\lesssim1/\Gamma$ of the shell 
\citep[Eq.~5.55 of][]{Dermer2009},
\begin{equation}
\Delta t_{\Delta,\rm obs} \simeq \frac{(1+z)\,\Delta}{\Gamma c},
\qquad
\Delta t_{\rm cur, obs} \simeq \frac{(1+z)\,r}{2\Gamma^2 c}.
\label{eq:timescales}
\end{equation}
We consider the case in which the two are equal and set the observed variability timescale, $\Delta t_{\Delta,\rm obs}= \Delta t_{\rm cur, obs}=\tvar$, so that
\begin{equation}
\Delta = \frac{\Gamma c\,\tvar}{1+z},
\qquad
r = 2\Gamma\Delta = \frac{2\Gamma^2 c\,\tvar}{1+z}.
\label{eq:shell_sizes}
\end{equation}
Then, the shell volume is $V_{\rm sh} = 4\pi r^2 \Delta$. If the burst
consists of $N$ such shells (pulses), the total isotropic-equivalent energy
emitted over the burst duration $T_{\rm obs}\simeq N\,\tvar$ is
$E_{\gamma,\rm iso}=L_{\gamma,\rm iso}\,T_{\rm obs}/(1+z)$. The $\gamma$-ray
energy radiated per shell, measured in the shell rest frame, is
$E_\gamma=E_{\gamma,\rm iso}/(\Gamma N)$, so that the radiation energy density in the shell frame is
\begin{equation}
u_{\rm sh} = \frac{E_\gamma}{V_{\rm sh}}
= \frac{L_{\gamma,\rm iso}}{4\pi r^2\, c\, \Gamma^2}
= \frac{(1+z)^2\, L_{\gamma,\rm iso}}{16\pi\,\Gamma^6 c^3\,\tvar^2}
\label{eq:u_shell}
\end{equation}
which the second expression is the standard one for the energy density in the rest frame of the GRB emitting region.

Comparing Eqs.~\ref{eq:u_shell} and \ref{eq:u_blob}, which are both
expressed in terms of the same observables, $L_{\gamma,\rm iso}$ and $\tvar$,
we obtain
\begin{equation}
\frac{u_{\rm sh}}{u_{\rm b}} = \frac{16}{3}.
\label{eq:ratio_shell}
\end{equation}
The origin of this factor is geometrical. Emission from the shell reaches the
observer only from the region $\theta\lesssim 1/\Gamma$, whose transverse size,
measured in the shell rest frame, is $r/\Gamma = 2\Delta$. This region is
causally connected, and is effectively the blob. The
causality-limited blob of Eq.~\ref{eq:R_blob}, however, has the maximum allowed radius
$R = 2\Gamma c\,\tvar/(1+z) = 2\Delta = r/\Gamma$. Its diameter is twice the size of the visible patch, and its energy density is correspondingly lower. Had we instead identified the blob diameter with the visible patch,
$2R=r/\Gamma$ (i.e.\ $R=\Delta$), Eq.~\ref{eq:u_blob} would give a value
four times larger, and the ratio would be $u_{\rm sh}/u_{\rm b}=4/3 \sim 1$. 

The different dependence of the received flux on the bulk motion,
$\propto \delta ^4$ for a blob versus $\propto\Gamma^2$ for a shell, reflects the fact that the luminosity of the
whole shell is the sum over $\sim\Gamma^2$ such patches, only one of which is
directed towards the observer; the energy density within each patch is the same.

The blob of Eq.~\ref{eq:R_blob} is the largest region compatible with the
observed variability and therefore gives the minimum energy density. The
consistent counterpart is the shell configuration that likewise minimises the
energy density for a given variability: a thin shell, $\Delta\ll r/\Gamma$,
located at the largest radius allowed by the curvature timescale, discussed
next.

\citet[][Chapter 5.5.3]{Dermer2009} computed the light curve of a thin shell,
$\Delta t_{\Delta, \rm obs}/ \Delta t_{\rm cur, obs} \ll 1$ (or equivalently $\Delta\ll r/\Gamma$), that radiates for only a short time, so that the pulse is
shaped by the arrival delay of high-latitude emission (the curvature limit).
Relating $\tvar$ with the full-width-at-half-maximum (FWHM) of this pulse, they showed that the shell radius is
$r\simeq7.7\,\Gamma^2c\,\tvar/(1+z)$, a factor $\simeq3.8$ further out than in
Eq.~\ref{eq:shell_sizes}. If the shell radiates for a fraction $\eta_t$ of
the curvature timescale, its energy density, in bolometric form, becomes
\begin{equation}
u_{\rm sh,cur}=\frac{(2^{1/3}-1)^2\,(1+z)^2\,L_{\gamma,\rm iso}}
{64\pi\,\eta_t\,\Gamma^6 c^3\,\tvar^2}.
\end{equation}
Therefore, the ratio of energy densities reads
\begin{equation}
\frac{u_{\rm sh,cur}}{u_{\rm b}}=\frac{4\,(2^{1/3}-1)^2}{3\,\eta_t}\approx0.9 \quad({\rm for} \quad \eta_t\approx0.1).
\label{eq:ratio_DM}
\end{equation}

Summarizing, for a given luminosity and variability timescale, the minimum energy density of a causality-limited blob, $R =\delta c\,\tvar/(1+z)$, agrees with that
of a curvature-limited relativistic shell to within a factor of order unity
(Eq.~\ref{eq:ratio_DM}). For a shell whose radial and angular timescales are
equal, the energy density is a factor $16/3$ higher than that of the
causality-limited blob (Eq.~\ref{eq:ratio_shell}); this difference reflects only the size of the blob relative to the visible, causally connected patch of
the shell, and reduces to $4/3$ when the two are identified. The results are therefore not sensitive to the assumed geometry for an on-axis observer of a uniform jet.

\section{Updates to LeHaMoC and numerical validation}\label{app:code-updates}
We summarize the updates implemented in the version of \code used in this work and describe the numerical checks relevant for the calculations presented in the main text. The original version of \code is a time-dependent leptohadronic radiative-transfer code that solves the coupled kinetic equations for relativistic particles and photons in a homogeneous magnetized source. The version used here contains the same general numerical framework, but includes several updates that improve its flexibility and efficiency and that are particularly relevant for prompt-emission applications.

\subsection{Synchrotron and inverse-Compton cooling}
The synchrotron and inverse Compton (IC) loss terms have been updated so that the ultra-relativistic factor $\gamma^2$ is replaced by $\beta^2\gamma^2=\gamma^2-1$. This correction is negligible for the highly relativistic particles, but ensures the physically correct behavior of the loss rate as $\gamma\rightarrow1$. The IC cooling rate is computed including Klein-Nishina suppression. We use the approximation of \cite{2005MNRAS.363..954M}, in which the cooling rate for a single electron due to IC scattering with an isotropic photon field can be approximated by 
\begin{equation}
    -\frac{{\rm d}\gamma_{\rm IC}}{{\rm d}t}=\frac{4}{3} \frac{\sigma_T c}{m_ec^2} \gamma^2 \int {\rm d}\epsilon\, u_{\rm ph}(\epsilon)F_{\rm KN}\left(4\gamma\epsilon\right),
    \label{Eq:dg_dt_ic}
\end{equation}
where $\epsilon=h \nu/(m_{\rm e}c^2)$, $u_{\rm ph}(\epsilon)$ is the differential energy density of photons, and $F_{\rm KN}\left(4\gamma\epsilon\right)$ is the Klein-Nishina suppression factor \citep[see Eq.~A12 in][]{2005MNRAS.363..954M}. For $4\gamma \epsilon \ll 1$, $F_{\rm KN}\rightarrow1$, and Eq.~\ref{Eq:dg_dt_ic} reduces to the standard Thomson cooling rate. 

As a validation test of the implementation of the Klein-Nishina cooling module, we compare the steady-state electron distribution obtained with \code with the analytical asymptotic solution of   \citep{1970RvMP...42..237B} (see Eq.~5.25). We consider a homogeneous spherical source of radius $R=10^{15}$~cm filled with photons of characteristic frequency $\nu_0=10^{16.4}$~Hz and total energy density $u_{\rm ph}=2\cdot 10^5$~$\rm erg\, cm^{-3}$. For this photon field, the transition to the Klein-Nishina regime occurs at $\gamma_{\rm KN}\simeq\frac{m_ec^2}{h\nu_0}\simeq 5\cdot10^3$. We therefore inject electrons only above this Lorentz factor, using a power-law distribution between $\gamma_{\rm min}=10^4$ and $\gamma_{\rm max}=10^7$. The injected luminosity is $L_{\rm e}=10^{40}$~$\rm erg\, s^{-1}$, and we repeat the calculation for injection indices $p_{\rm el}=1.5,\,2$, and $3$. All other loss processes, including electron escape, are disabled, so that the electron distribution is shaped only by IC cooling in the Klein–Nishina regime. 

\begin{figure}
    \centering
    \includegraphics[width=0.95\linewidth]{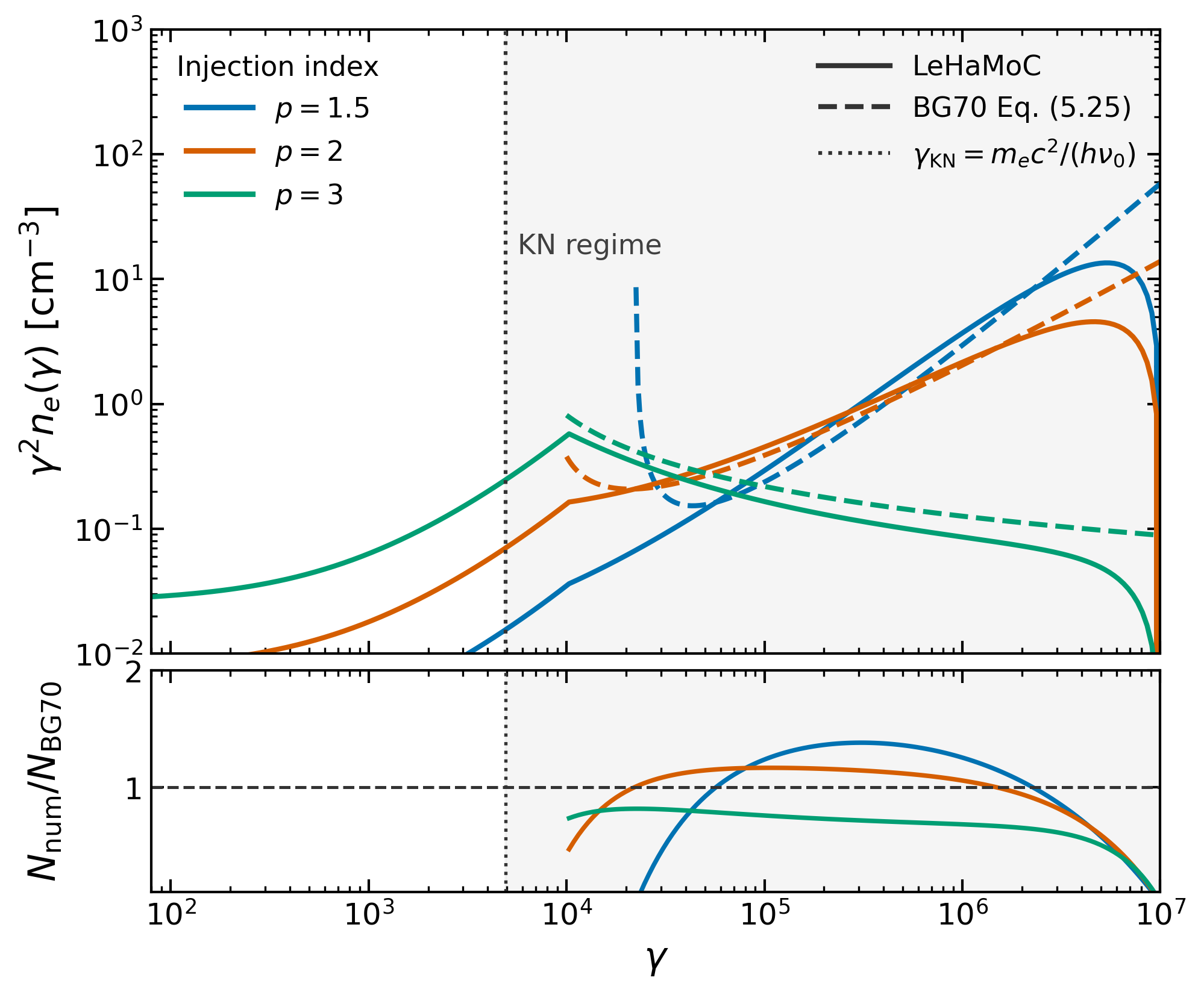}
    \caption{Validation of the Klein-Nishina cooling implementation for different injection indices of the electron/positron distribution. The upper panel compares the steady-state electron distributions computed with \code to the asymptotic Klein--Nishina solutions of \citet{1970RvMP...42..237B} for $p=1.5$, $2$, and $3$. Solid lines show the numerical results, while dashed lines show the analytical BG70 solutions. The vertical dotted line marks $\gamma_{\rm KN}=m_ec^2/(h\nu_0)$, and the lower panel shows the ratio between the numerical and analytical solutions.}
    \label{fig:analytical_num_kn}
\end{figure}

In Fig.~\ref{fig:analytical_num_kn} we compare the steady-state electron distributions computed with \code to the analytical asymptotic solutions of \citet{1970RvMP...42..237B}. The numerical solution reproduces the overall shape of the analytical distribution for $\gamma\gtrsim10^4$, where the injected electrons are above the characteristic Klein--Nishina transition scale. Across the interval $10^4\lesssim\gamma\lesssim10^6$, the agreement is within a factor of order unity, with deviations typically smaller than about 50\%. This comparison should be interpreted as a validation of the asymptotic scaling and normalization of the Klein-Nishina cooling treatment, since the analytical expression of \citet{1970RvMP...42..237B} is derived for an idealized extreme Klein-Nishina limit, whereas the numerical calculation uses the continuous-loss implementation adopted in \code. 

\subsection{Numerical acceleration}
Several numerically intensive functions have been accelerated using {\tt numba.njit} \citep{2015llvm.confE...1L}. The just-in-time compilation is used only to reduce the runtime of loop operations, such as radiative kernel calculations that demand energy grid integrations. It does not change the numerical method or the physical approximations. The updated code also contains tabulated interpolation functions for the photomeson interactions and the production of the secondary particles including photon, electron/positron, and neutrino distributions based on the interaction kernels described in \citep{2008PhRvD..78c4013K}. A SSC model evaluation
requires approximately 0.2$-$0.3~s and a leptohadronic evaluation requires $\sim10$~s on an Apple M4 central processing unit of a MacBook Air laptop.

\section{Cooling timescales}\label{app:cool} 

\begin{figure*}
    \centering
    \includegraphics[width=0.9\textwidth]{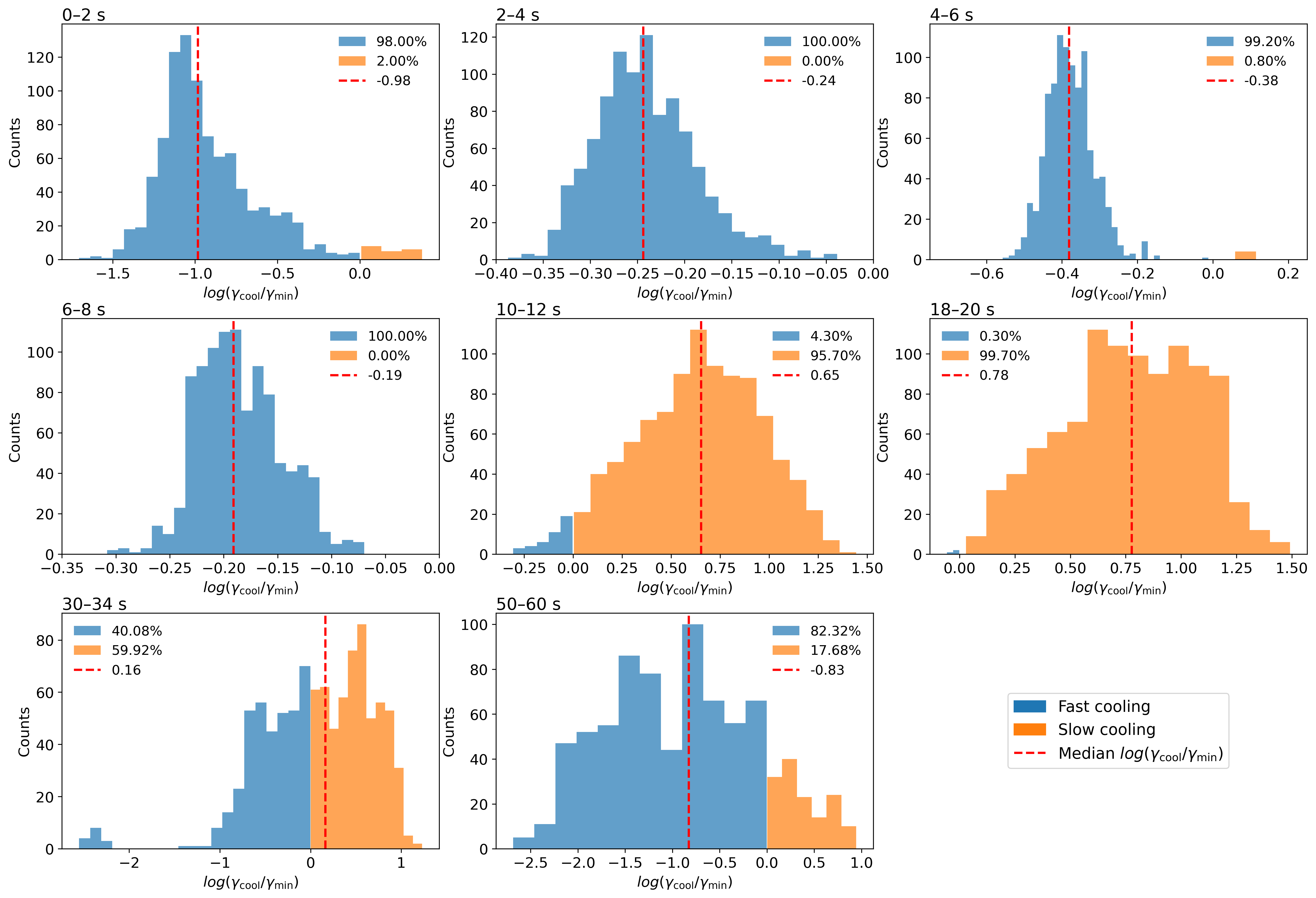}
    \caption{Histograms of $\log_{10}(\gamma_{\rm cool}/\gamma_{\min})$ for the eight time intervals under study. The values of $\log_{10}\left(\gamma_{\rm cool}/\gamma_{\rm min}\right)$ are predominantly negative throughout the first four time intervals, indicating that the electron population is in the fast cooling regime. In the following two time intervals, a great increase of the percentage of positive values indicates that the electron population progressively transitions towards the slow cooling regime. In the last two time intervals the percentage of negative values increases again, indicating a transition towards the fast cooling regime once again.}
    \label{fig:gcool}
\end{figure*}

\begin{figure*}
    \centering
    \includegraphics[width=0.9\textwidth]{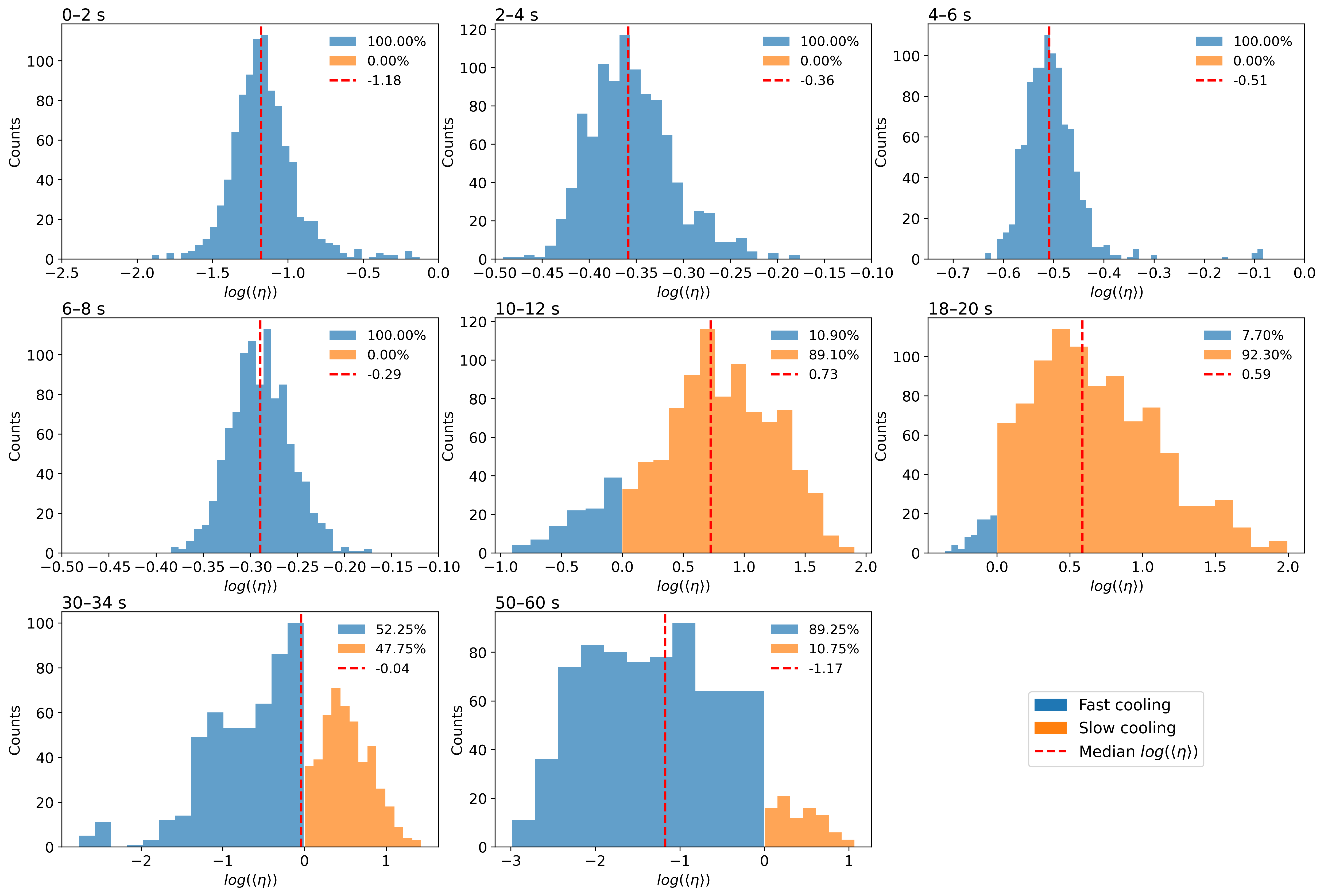}
    \caption{Histograms of $\log_{10}\left(\langle \eta \rangle\right)$ for the eight time intervals under study. The values of $\log_{10}\left(\langle \eta \rangle\right)$ are predominantly negative throughout the first four time intervals, indicating that the system remains in the fast cooling regime. In the following two time intervals, the percentage of positive values increases, pointing that the electron population progressively transitions towards the slow cooling regime. The last two time intervals showcast a transition towards the fast cooling regime, since the percentage of positive values decreases again.}
    \label{fig:average_eta}
\end{figure*}

We calculate the particle cooling timescales of the dominant radiative processes in the emitting region to evaluate the cooling regimes for the time intervals presented in Section~\ref{sec:results}. 

The synchrotron energy loss timescale is given by

\begin{equation}
    t_{\rm cool, syn} = \frac{E}{\left(-{\rm d}E/{\rm d}t\right)}_{\rm syn} = \frac{3 m_{e}c}{4\sigma_{T} \gamma \, u_{\rm B}},
    \label{eq:t_syn}
\end{equation}
where $u_B = B^2/(8\pi)$. Similarly, the inverse Compton cooling timescale is written as

\begin{equation}
    t_{\rm cool, ics} = \frac{E}{\left(-{\rm d}E/{\rm d}t\right)}_{\rm ics} = \frac{3m_{e}c}{4 \sigma_T \gamma} \frac{1}{\int d\epsilon\, u_{\rm ph}(\epsilon)\, F_{\rm KN}\left(4\gamma\epsilon\right)},
    \label{eq:t_ics}
\end{equation}
where $u_{\rm ph}(\epsilon)$ denotes the differential energy density of the photons, $\epsilon=h \nu/(m_{\rm e}c^2)$ and $F_{\rm KN}\left(4\gamma\epsilon\right)$ is a function that approximates the energy losses in the Klein-Nishina regime (see also Appendix~\ref{app:code-updates}). 

The total radiative cooling timescale is then obtained by combining the individual contributions as

\begin{equation}
    t^{-1}_{\rm cool} = t^{-1}_{\rm cool, syn} + t^{-1}_{\rm cool, ics}.
    \label{eq:t_cool}
\end{equation}

Using 1000 randomly selected samples from the posterior distributions, we calculate the cooling Lorentz factor of the electron distribution, which is defined as $t_{\rm cool}(\gamma_{\rm cool}) = t_{\rm esc} = R/c$. In Fig.~\ref{fig:gcool} we present histograms of $\log_{10}(\gamma_{\rm cool}/\gamma_{\min})$ for each time interval. We observe that during the early emission stage, throughout the first four time windows, the values of $\log_{10}\left(\gamma_{\rm cool}/\gamma_{\rm min}\right)$ are predominantly negative. Since $\gamma_{\rm cool} < \gamma_{\rm min}$ is the defining condition for the fast cooling regime, these negative logarithmic values indicate that the emitting electron population remains in the fast cooling regime throughout this phase.
On the contrary, during the late emission stage, the percentage of positive logarithmic values increases substantially. Since positive values imply $\gamma_{\rm min} < \gamma_{\rm cool}$, they suggest a transition of the radiating particle population from the fast to the slow cooling regime.

An alternative way of evaluating the radiative efficiency of the electron distribution is through the distribution-weighted average of the ratio $\eta(\gamma) = t_{\rm cool} \left(\gamma\right)/t_{\rm esc}$, namely

\begin{equation}
    \langle \eta \rangle = \frac{\int {\rm d}\gamma\, \frac{{\rm d}N}{{\rm d}\gamma}\, \eta\left(\gamma\right)}{\int {\rm d}\gamma\, \frac{{\rm d}N}{{\rm d}\gamma}},
    \label{eq:avg_eta}
\end{equation}
where $\frac{{\rm d}N}{{\rm d}\gamma} \propto \gamma^{-p}$ is the distribution of the injected electrons. If $\langle \eta \rangle \gg 1$, particles generally escape before cooling, which indicates that the system is in the slow cooling regime. Consequently, positive $\log_{10}\left(\langle \eta \rangle\right)$ values correspond to slow cooling, whereas negative ones to fast cooling.

The histograms of $\log_{10}\left(\langle \eta \rangle\right)$ are shown in Fig.~\ref{fig:average_eta} for all time windows. They reveal a similar picture qualitatively to those of Fig.~\ref{fig:gcool}, regarding the determination of the fast and slow cooling regimes. \citet{Gompertz2023} report that the spectral photon indices are consistent with a fast cooling regime up to $\sim42$~s, extending well beyond the first four time intervals in our analysis (up to $\sim$8~s), which clearly favor fast-cooling solutions.
It is worth noting that in Fig.~3 of \citet{Gompertz2023}, the photon indices begin to show a mild deviation from the expected fast cooling values after 10$-$20~s. 


\section{Relativistic blobs within the jet: average properties}\label{app:minijet}
In this appendix, we adopt the framework presented in Section~\ref{sec:minijet} and look for combinations of $\sigma, \theta_{\rm co}$ that can reproduce the Doppler factor values derived in Section~\ref{sec:results} for the different time intervals.
  
 \begin{figure*}
\includegraphics[width = 0.33\textwidth]{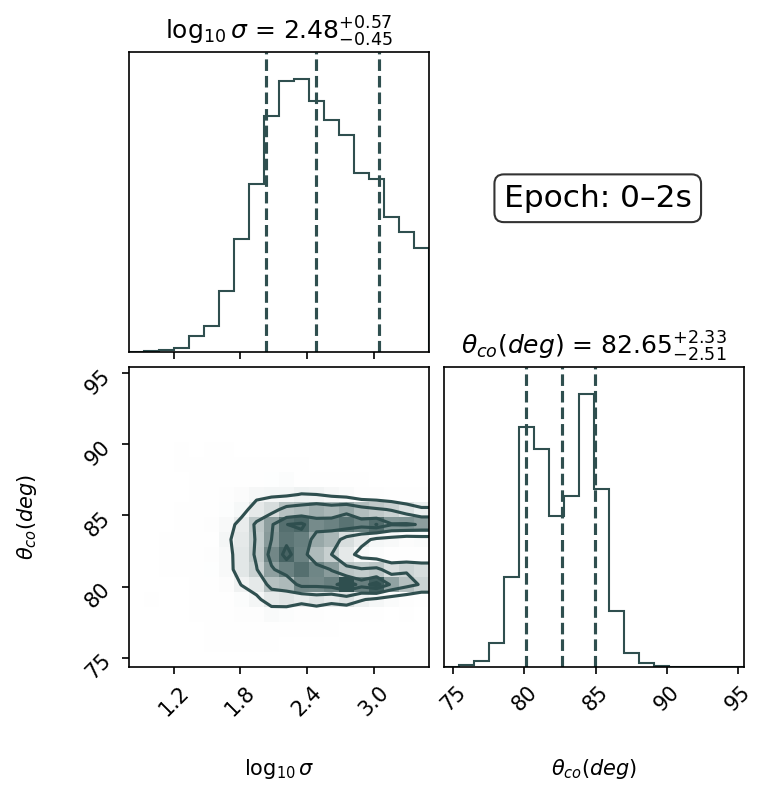}   
\includegraphics[width = 0.33\textwidth]{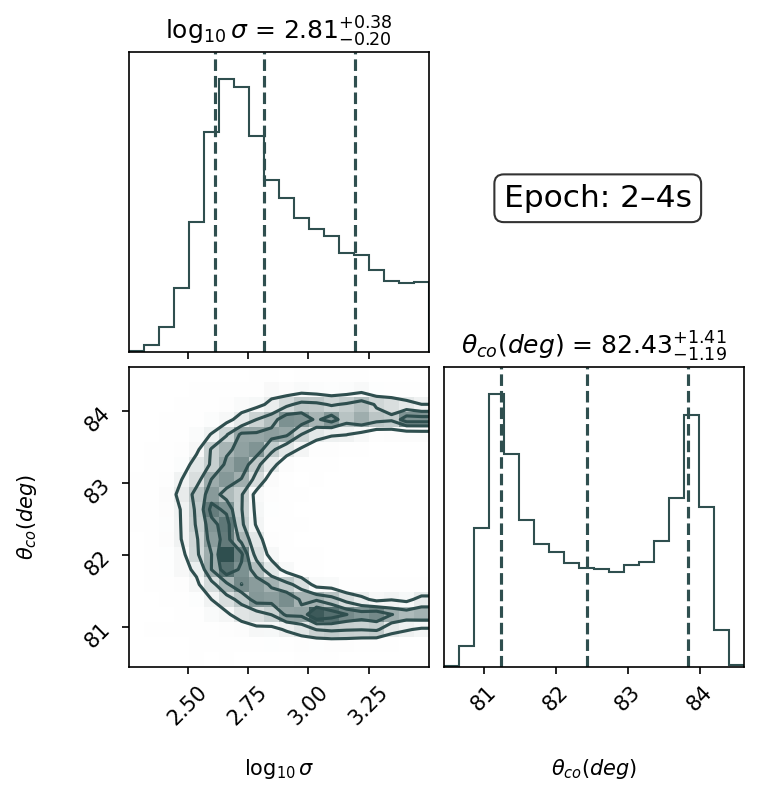}   
\includegraphics[width = 0.33\textwidth]{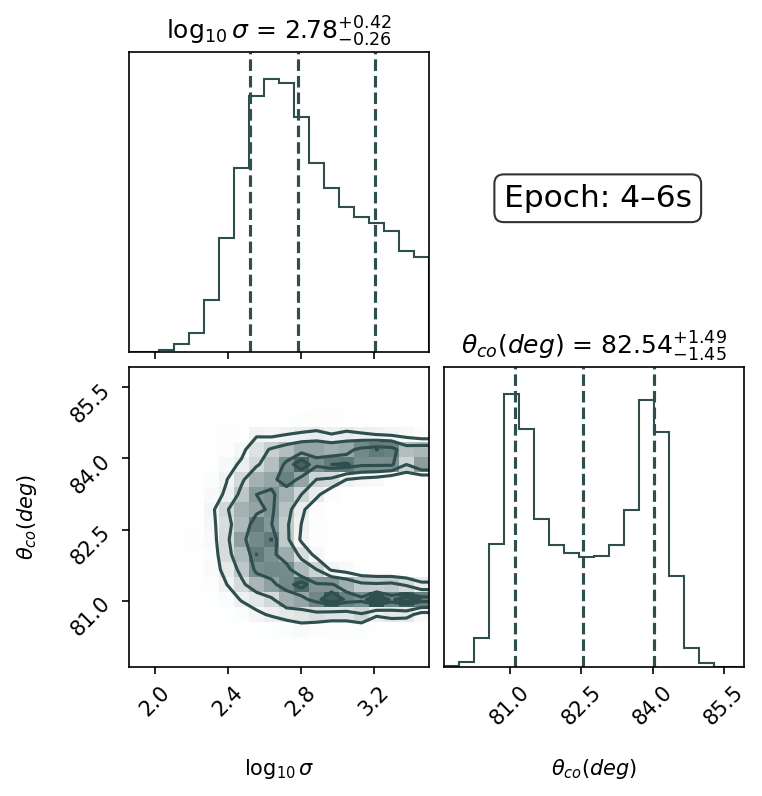}   
\includegraphics[width = 0.33\textwidth]{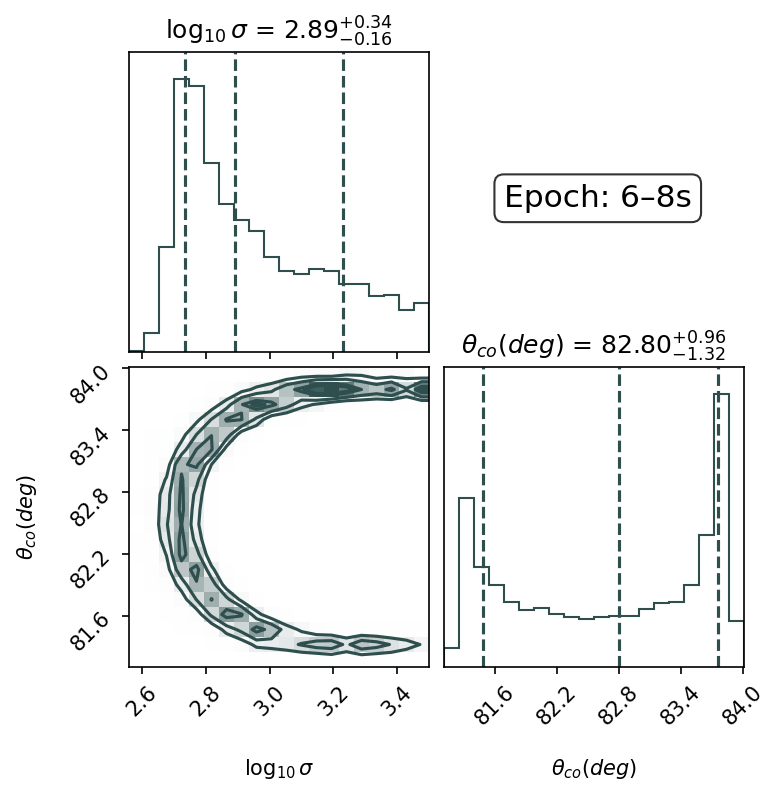}   
\includegraphics[width = 0.33\textwidth]{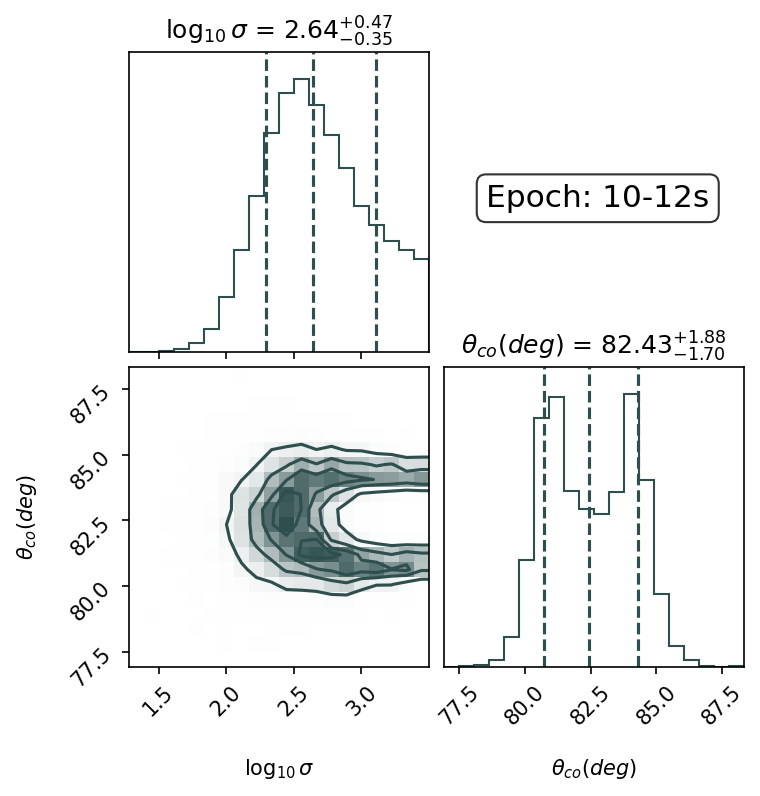}   
\includegraphics[width = 0.33\textwidth]{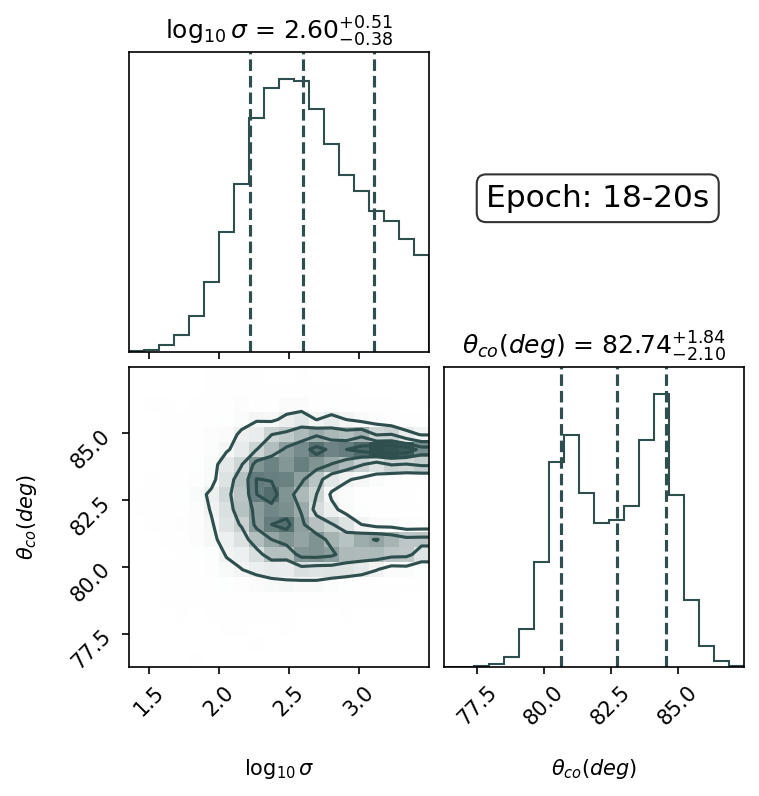}   
\includegraphics[width = 0.33\textwidth]{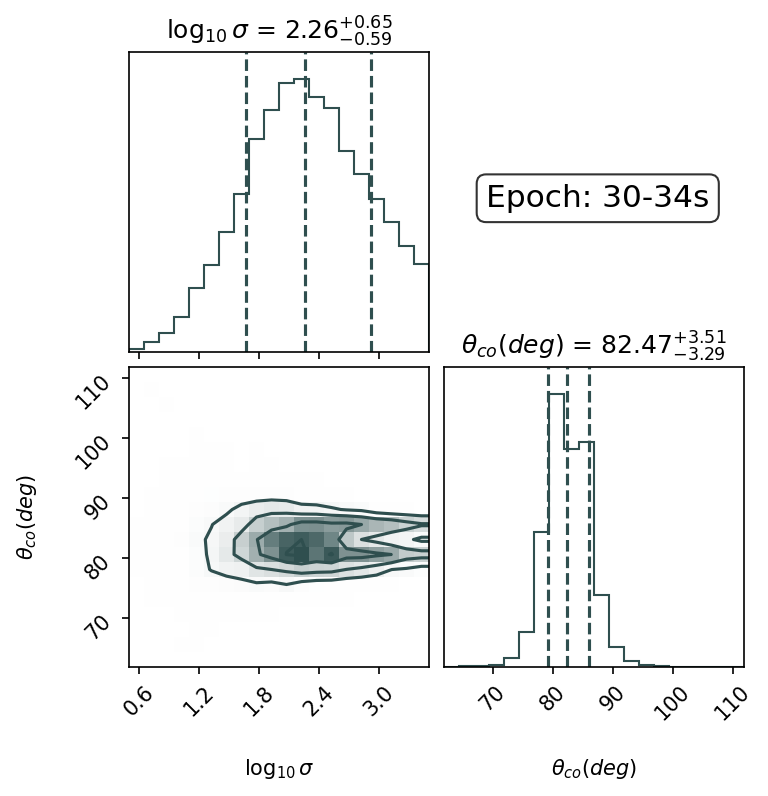}   
\includegraphics[width = 0.33\textwidth]{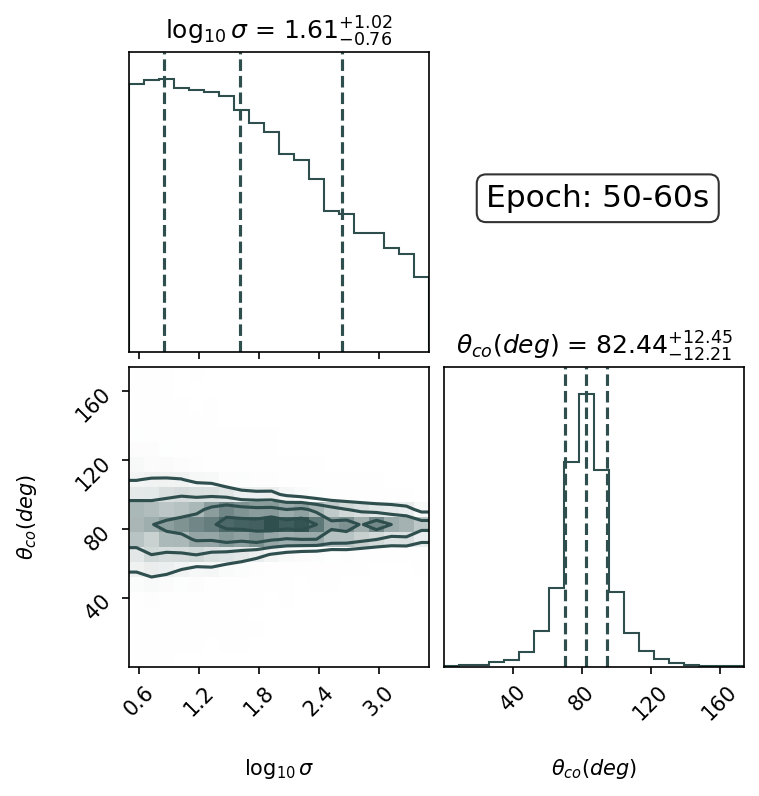} 
\caption{Corner plots showing the posterior distributions of $\theta_{\rm co}$ and $\sigma$ for different time windows.}
\label{fig:corner-blob}
 \end{figure*}
 
For this purpose, we adopt the median value and the 68\% credible regions derived for $\delta$ from the posterior distributions shown in Fig.~\ref{fig:violin} and perform a MCMC fit to the parameters $\sigma, \theta_{\rm co}$ for fixed jet Lorentz factor $\Gamma_j = 73.11$ and viewing angle $\theta_{\rm obs} = 0.688$~deg. The resulting posterior distributions are presented  in Fig.~\ref{fig:corner-blob}. We infer jet plasma magnetizations in the range $\sigma \sim$ 40$-$800 with the lower value corresponding to the last time interval, where the emission is softer and dimmer. Blobs moving at large angles ($\theta_{\rm co}\sim 80$~deg) with respect to the jet axis result in the largest Doppler boosting, as also shown in Fig.~\ref{fig:heatmap} for an indicative value of $\sigma = 100$. The large values of $\theta_{\rm co}$ translate to $\theta \sim 1/\Gamma_{\rm j}$ in the observer's frame. In other words, the plasmoid direction of motion is aligned with the line of sight in the observer's frame.

As discussed in Section~\ref{sec:discussion}, the inferred values of $\sigma$ and $\theta_{\rm co}$ should be interpreted as effective, time-averaged properties of the population of emitting regions contributing to the emission within each time interval, rather than as the properties of individual blobs. A more realistic treatment would require time-resolved spectroscopy of individual pulses, allowing the physical properties of each emitting region to be inferred separately within this physical framework.

\end{document}